\documentclass[preprint, 12pt]{elsarticle}
\biboptions{sort&compress}

\usepackage{amssymb}

\usepackage{xcolor} % for color.
\usepackage[a4paper]{geometry}
\usepackage[hidelinks]{hyperref} % for hyperlink.
\usepackage{multirow}
\usepackage{physics}
\usepackage{slashed}
\usepackage{subcaption}
\usepackage{tabularx}
\usepackage{ulem}
    \renewcommand{\sout}{\bgroup \color{red} \ULdepth=-.5ex \ULset}

\newcommand{\mathii}{\mathrm{i}}
\newcommand{\ieta}{\mathii\eta}
\newcommand{\orcidlogo}{\includegraphics[height=\fontcharht\font`A]{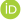}}
\newcommand{\orcid}[1]{\href{https://orcid.org/#1}{\textcolor[HTML]{A6CE39}{\orcidlogo}}}

\newcounter{bla}
\newenvironment{refnummer}{
    \list{[\arabic{bla}]}{
        \usecounter{bla}
        \setlength{\itemindent}{0pt}
        \setlength{\topsep}{0pt}
        \setlength{\itemsep}{0pt}
        \setlength{\labelsep}{2pt}
        \setlength{\listparindent}{0pt}
        \settowidth{\labelwidth}{[9]}
        \setlength{\leftmargin}{\labelwidth}
        \addtolength{\leftmargin}{\labelsep}
        \setlength{\rightmargin}{0pt}
    }
}{\endlist}

\journal{Computer Physics Communications}

\begin{document}

\begin{frontmatter}



\title{\textsc{FeAmGen.jl}: A Julia Program for \textbf{Fe}ynman \textbf{Am}plitude \textbf{Gen}eration}


\author[IHEP,UCAS]{Quan-feng Wu}\ead{wuquanfeng@ihep.ac.cn}
\author[IHEP,UCAS,CHEP]{Zhao Li\corref{main}}\ead{zhaoli@ihep.ac.cn}

\cortext[main]{Corresponding author.}

\address[IHEP]{Institute of High Energy Physics, Chinese Academy of Sciences, Beijing 100049, China}
\address[UCAS]{University of Chinese Academy of Sciences, Beijing 100049, China}
\address[CHEP]{Center for High Energy Physics, Peking University, Beijing 100871, China}

% FeAmGen-Release (c) by Quan-feng WU <wuquanfeng@ihep.ac.cn> and
% Zhao Li <zhaoli@ihep.ac.cn>
% 
% FeAmGen-Release is licensed under a
% Creative Commons Attribution 4.0 International License.
% 
% You should have received a copy of the license along with this
% work. If not, see <http://creativecommons.org/licenses/by/4.0/>.

\begin{abstract}
    \textsc{FeAmGen.jl} is a Julia package designed to generate Feynman diagrams and their corresponding amplitudes for various processes in particle physics.
    Utilizing the models in the Universal Feynman Output (UFO) format and \textsc{Qgraf} for diagram generation, it also employs \textsc{SymEngine.jl} and \textsc{Form} for amplitude generation.
    Additionally, the package offers functions for constructing Feynman integral topologies.
    In conclusion, \textsc{FeAmGen.jl} provides usability and versatility for the high precision calculations of the perturbative quantum field theory in the Standard Model or in the extensions beyond it.
    The corresponding codes are available at \url{https://code.ihep.ac.cn/IHEP-Multiloop/FeAmGen.jl.git}.
    %This paper provides an introduction to \textsc{FeAmGen.jl} in detail.
\end{abstract}


\end{frontmatter}

% \linenumbers

\newpage

\noindent
\textbf{PROGRAM SUMMARY}
\vspace{1cm}

\begin{small}
    \noindent
    \emph{Program Title:} \textsc{FeAmGen.jl} \\
    \emph{CPC Library link to program files:} (to be added by Technical Editor) \\
    \emph{Developer's repository link:} \url{https://code.ihep.ac.cn/IHEP-Multiloop/FeAmGen.jl} \\
    \emph{Code Ocean capsule:} (to be added by Technical Editor) \\
    \emph{Licensing provisions:} \href{https://mit-license.org}{MIT License} \\
    \emph{Programming language:} \hyperlink{https://julialang.org/}{The Julia Programming Language} \\
    \emph{External routines/libraries used:} \begin{enumerate}
        \item Julia Standard Libraries: \textsc{Dates} [1], \textsc{Pkg} [2];
        \item Julia Packages: \textsc{AbstractAlgebra.jl} [3], \textsc{Combinatorics.jl} [4], \textsc{FORM\_jll.jl} [5], \textsc{JLD2.jl} [6], \textsc{nauty\_jll.jl} [7], \textsc{OrderedCollections.jl} [8], \textsc{Pipe.jl} [9], \textsc{PyCall.jl} [10], \textsc{SymEngine.jl} [11], \textsc{YAML.jl} [12];
        \item Non-Julia Programs/Packages: \textsc{Form} [13], \textsc{nauty} [14], \textsc{Qgraf} [15], \textsc{SymEngine} [16], The \textsc{Python} Programming Language [17].
    \end{enumerate}
    \emph{Nature of problem:} The generation of the Feynman diagrams and the corresponding amplitudes in the quantum field theory. \\ % Describe the nature of the problem here.
    \emph{Solution method:} We present the Julia package \textsc{FeAmGen.jl}, which allows users to define arbitrary models and couplings conveniently in the Universal Feynman Output (UFO) format and to call \textsc{Qgraf} directly to generate Feynman diagrams and amplitudes for arbitrary processes. Some symbolic simplifications will be applied to the amplitudes automatically by calling \textsc{Form}, the well-known computer algebra system for big expressions. The diagrams are written in the \LaTeX{} files that utilize \textsc{TikZ-Feynman} for creating beautiful and clear figures. Finally, we also provide the algorithm to construct Feynman integral topologies.\\ % Describe the method solution here.
    \emph{Additional comments:} Thanks to the Julia Yggdrasil Project, this package can run on almost all operating systems except Windows, where WSL2 is recommended. Other restrictions are determined by the CPU time and the available RAM of the computer. \\ %Provide any additional comments here.
    \emph{References:}
        \begin{refnummer}
            \item \textit{Dates $\cdot$ The Julia Language}. URL: \url{https://docs.julialang.org/en/v1/stdlib/Dates}.
            \item \textit{Pkg $\cdot$ The Julia Language}. URL: \url{https://docs.julialang.org/en/v1/stdlib/Pkg}.
            \item \textit{AbstractAlgebra.jl}. URL: \url{https://nemocas.github.io/AbstractAlgebra.jl}.
            \item \textit{Combinatorics.jl}. URL: \url{https://juliamath.github.io/Combinatorics.jl}.
            \item \textit{JuliaBinaryWrappers/FORM\_jll.jl}. URL: \url{https://github.com/JuliaBinaryWrappers/FORM_jll.jl}.
            \item \textit{Julia Data Format - JLD2}. URL: \url{https://juliaio.github.io/JLD2.jl/stable}.
            \item \textit{JuliaBinaryWrappers/nauty\_jll.jl}. URL: \url{https://github.com/JuliaBinaryWrappers/nauty_jll.jl}.
            \item \textit{OrderedCollections.jl}. URL: \url{https://juliacollections.github.io/OrderedCollections.jl/latest}.
            \item \textit{oxinabox/Pipe.jl: An enhancement to julia piping syntax}. URL: \url{https://github.com/oxinabox/Pipe.jl}.
            \item \textit{JuliaPy/PyCall.jl: Package to call Python functions from the Julia language}. URL: \url{https://github.com/JuliaPy/PyCall.jl}.
            \item \textit{SymEngine.jl: Julia wrappers of SymEngine}. URL: \url{https://symengine.org/SymEngine.jl}.
            \item \textit{JuliaData/YAML.jl: Parse yer YAMLs}. URL: \url{https://github.com/JuliaData/YAML.jl}.
            \item \textit{The FORM project for symbolic manipulation of very big expressions}. URL: \url{https://www.nikhef.nl/~form/}.
            \item \textit{nauty and Traces: graph canonical labeling and automorphism group computation}. URL: \url{https://pallini.di.uniroma1.it}.
            \item \textit{Qgraf — a computer program that generates Feynman diagrams for various types of QFT models}. URL: \url{http://cfif.ist.utl.pt/~paulo/qgraf.html}.
            \item \textit{SymEngine: a fast symbolic manipulation library written in C++} \url{https://github.com/symengine/symengine}
            \item \textit{Python Programming Language}. URL: \url{https://www.python.org}.
        \end{refnummer}
\end{small}

\newpage

% FeAmGen-Release (c) by Quan-feng WU <wuquanfeng@ihep.ac.cn> and
% Zhao Li <zhaoli@ihep.ac.cn>
% 
% FeAmGen-Release is licensed under a
% Creative Commons Attribution 4.0 International License.
% 
% You should have received a copy of the license along with this
% work. If not, see <http://creativecommons.org/licenses/by/4.0/>.

\section{Introduction}
\label{sec:intro}

The quantum field theory for particle physics is very successful because of the stunning agreements between the predictions of theory and the data from experiments. 
In particular, the discovery of the Higgs boson at the CERN Large Hadron Collider (LHC) in 2012 provided the final puzzle piece to the Standard Model (SM) \cite{CMS:2012qbp, ATLAS:2012yve}. 
Since then, the high precision investigations into the properties of the Higgs boson and the other SM particles have become the primary interest within the high-energy physics community, necessitating more advanced experiment data.
Several future projects have been proposed to meet this demand in recent years. 
The High-Luminosity LHC \cite{Apollinari:2015wtw} was proposed to increase the integrated luminosity by a factor of 10 beyond the design value of the present LHC project so that the statistical error can be reduced significantly. Furthermore, to achieve the most precise investigation on the Higgs boson and other sensitive observables, several designs of the Higgs factory were proposed, including the International Linear Collider (ILC) \cite{ILC:2013jhg, Behnke:2013xla, Bambade:2019fyw}, the Circular Electron Positron Collider (CEPC) \cite{CEPCStudyGroup:2018ghi, CEPCStudyGroup:2018rmc}, and the Future Circular Collider (FCC) \cite{TLEPDesignStudyWorkingGroup:2013myl, FCC:2018byv, FCC:2018evy}.
With the steady improvement of experiment accuracy, the particle physics community urgently needs matching theoretical predictions, where the calculations of the higher-loop Feynman diagrams are necessary.

The Feynman diagram technique is the core of the perturbative quantum field theory (pQFT), where the theoretical calculation relies on generating the Feynman diagrams and the corresponding amplitude expressions.
However, the number of corresponding Feynman diagrams could increase exponentially for the high-order corrections to the scattering processes \cite{Veltman:1994wz, Smirnov:2012gma, Henn:2014yza, Elvang:2015rqa, Li:2020ign, Weinzierl:2022eaz}.
Furthermore, several algebraic manipulations, e.g., the color algebra, the Dirac algebra, etc., should be applied.
Reducing a large number of Feynman integrals into several master integrals is also necessary to improve the efficiency of the calculation, where the integration-by-parts (IBP) techniques \cite{Chetyrkin:1981qh, Tkachov:1981wb} are applied.
Finally, the master integrals should be evaluated in a symbolic or numerical approach.
These works require huge computation resources, which makes it almost impossible to obtain high precision theoretical predictions manually.

Therefore, the automatic generation and calculation of the Feynman diagrams and the corresponding amplitudes are necessary for the high energy physics community.
At the tree-level and one-loop level, several computer programs have been developed, which can automatically perform the theoretical calculations from the Feynman diagrams to cross sections and decay rates, e.g., \textsc{MadGraph5\_aMC@NLO} \cite{Alwall:2014hca}, \textsc{OpenLoops} \cite{Buccioni:2019sur}, \textsc{GoSam} \cite{Cullen:2011ac, GoSam:2014iqq}, etc.
Although fully automatic computer programs are not available yet for higher loop calculations, several program packages have been developed, e.g. \textsc{Caravel} \cite{Abreu:2020xvt}, \textsc{pySecDec} \cite{Heinrich:2023til} and \textsc{AMFlow} \cite{Liu:2022chg},  \textsc{FeynArts} \cite{Kublbeck:1990xc, Hahn:2000kx}, \textsc{Qgraf} \cite{Nogueira:1991ex}, \textsc{FeynCalc} \cite{Mertig:1990an, Shtabovenko:2016sxi, Shtabovenko:2020gxv}, \textsc{Pacakge-X} \cite{Patel:2016fam}, \textsc{FormCalc} \cite{Hahn:2016ebn}, \textsc{tapir} \cite{Gerlach:2022qnc}, \textsc{ALibrary}\footnote{\url{https://magv.github.io/alibrary}}, etc, where the computer algebra systems (CAS) such as \textsc{Mathematica}\footnote{A commercial software available at \url{https://www.wolfram.com/mathematica}.} and \textsc{Form} \cite{Vermaseren:2000nd, Ueda:2020wqk} are widely used.
Especially, the prior choice for calculating multi-loop Feynman diagrams has always been \textsc{Qgraf}, e.g. the high precision theoretical predictions for the Higgs boson production at the LHC. However, \textsc{Qgraf} is not easy to be extended for the multi-loop calculation in the physics beyond SM, e.g. the SM effective theory (SMEFT).

In this paper, we present the package \textsc{FeAmGen.jl} developed in the \textsc{Julia} programming language, which is a flexible dynamic language, appropriate for scientific and numerical computing, with performance comparable to traditional statically-typed languages\footnote{\url{https://julialang.org}}.
Recently, the \textsc{Julia} programming language has been used to develop several packages for high energy physics\footnote{\url{https://github.com/JuliaHEP}}.
Some investigations on its potential for high energy physics computing are reported in Refs.~\cite{Stanitzki:2020bnx, Eschle:2023ikn}, which show that the \textsc{Julia} is a promising programming language for high energy physics.
With the powerful package management system and the high performance provided by the \textsc{Julia} programming language, \textsc{FeAmGen.jl} could be integrated with the other packages developed in the \textsc{Julia} programming language to form a complete workflow for the automatic generation and evaluation of the multi-loop Feynman diagrams in the future.

\textsc{FeAmGen.jl} provides a user-friendly interface between \textsc{Qgraf} and the Universal FeynRules Output (UFO) format \cite{Degrande:2011ua}, which could be exported from the \textsc{Mathematica} package \textsc{FeynRules} \cite{Christensen:2009jx, Alloul:2013bka}. % Citation Darme:2023jdn for UFO v2 is removed.
Typically, the user provides the process information in the YAML format\footnote{\url{https://yaml.org}} and the theoretical model in the UFO format for \textsc{FeAmGen.jl}.
Then it can generate the Feynman diagrams via \textsc{Qgraf} and read the output of \textsc{Qgraf} to generate the amplitude expressions internally, where the symbolic manipulation is performed by \textsc{SymEngine.jl} and \textsc{Form} \cite{Ueda:2020wqk}.
Finally, the visualizations of the Feynman diagrams are presented in the \LaTeX{} format with the \textsc{TikZ-Feynman} package \cite{Ellis:2016jkw}.

For the calculation of higher loop Feynman diagrams, the identification and minimization of the Feynman topologies are also an important step, which has been implemented in several programs, e.g., \textsc{DIANA} \cite{Tentyukov:2002ig}, \textsc{tapir} \cite{Gerlach:2022qnc}, \textsc{q2e}/\textsc{exp} \cite{Harlander:1998cmq, Diaz:1999is}, \textsc{TopoID} \cite{Hoff:2016pot}, \textsc{feynson} \cite{Maheria:2022dsq}, \textsc{FeynCalc} in its version 10 \cite{Shtabovenko:2023idz}, etc.
\textsc{FeAmGen.jl} presents a canonical form of the Feynman integral, which helps to give a unique representation from the equivalent Feynman integrals.
The algorithm to minimize the number of corresponding Feynman topologies is also presented, which is based on the synthesis of Pak's algorithm \cite{Pak:2011xt, Wu:2023upw} and the canonical form of the Feynman integral as defined in this paper.

This paper is organized as follows.
The installation guide of \textsc{FeAmGen.jl} is provided in Sec.~\ref{sec:installation}.
In Sec.~\ref{sec:concept}, the basic concepts of the Feynman amplitude and the algorithms implemented in \textsc{FeAmGen.jl} are introduced.
The main functions provided by \textsc{FeAmGen.jl} are described in Sec.~\ref{sec:usage}, where several examples are also presented.
We make a summary in Sec.~\ref{sec:conclusion}.
In the \ref{apdx:output}, the output format of the amplitudes is described in detail.
\ref{apdx:kin} provides the definitions of the kinematic variables in \textsc{FeAmGen.jl}.
The algorithm for color algebra calculation is described in \ref{apdx:color}.
Pak's algorithm and its derivatives used in \textsc{FeAmGen.jl} are detailed in \ref{apdx:Pak}.

% FeAmGen-Release (c) by Quan-feng WU <wuquanfeng@ihep.ac.cn> and
% Zhao Li <zhaoli@ihep.ac.cn>
% 
% FeAmGen-Release is licensed under a
% Creative Commons Attribution 4.0 International License.
% 
% You should have received a copy of the license along with this
% work. If not, see <http://creativecommons.org/licenses/by/4.0/>.

\section{Installations}\label{sec:installation}

The convenient method to install \textsc{FeAmGen.jl} is to use the Julia package manager \textsc{Pkg.jl}, which requires \textsc{Julia} v1.6 or later.
The newcomers of \textsc{Julia} can refer to the official documentation\footnote{\url{https://docs.julialang.org}} for more information.
There are three main steps to install \textsc{FeAmGen.jl}:

\paragraph{Install \textsc{Julia}}
The \textsc{Julia} ecosystem can be installed whether using precompiled binaries or compiling from source by following the instructions at \url{https://julialang.org/downloads} directly.

\paragraph{Add IHEP-Multiloop Registry}
Julia registries contain information about packages, such as available releases and dependencies, and where they can be downloaded.
The General registry\footnote{\url{https://github.com/JuliaRegistries/General}} is the default registry maintained by the \textsc{Julia} community, which contains the main packages that are widely used in the \textsc{Julia} ecosystem.
The IHEP-Multiloop registry\footnote{\url{https://code.ihep.ac.cn/IHEP-Multiloop/JuliaRegistry}} provides the information about the package \textsc{FeAmGen.jl} and its dependencies developed and maintained by the authors.
The users should add the General registry and IHEP-Multiloop registry before installing \textsc{FeAmGen.jl} via the following commands:
\begin{center}
    \includegraphics{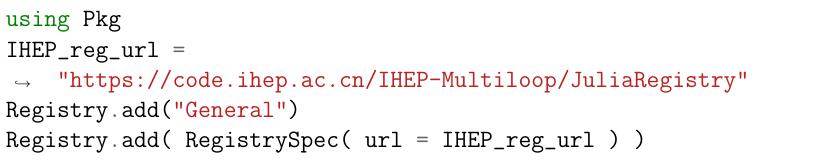}
\end{center}

\paragraph{Install \textsc{FeAmGen.jl}}
To install \textsc{FeAmGen.jl}, the users just need to execute the following commands:
\begin{center}
    \includegraphics{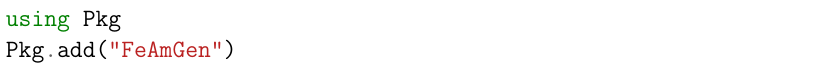}
\end{center}

\subsection{Additional Information}

\paragraph{External Non-\textsc{Julia} Programs}
\textsc{FeAmGen.jl} needs external non-\textsc{Julia} programs, \textsc{Form} and \textsc{Qgraf} as back-ends, which can be downloaded by \textsc{FeAmGen.jl} automatically\footnote{ At present \textsc{FeAmGen.jl} provides the automatic installation of \textsc{Form} v4.3.1 and \textsc{Qgraf} v3.6.7.}.
Besides, we also provide the option to use the external \textsc{Form} and \textsc{Qgraf} that have been installed locally.
This can be set by the environment variables \texttt{ENV["FORM"]} and \texttt{ENV["QGRAF"]}, which can be set in the \textsc{Julia} startup file\footnote{\url{https://docs.julialang.org/en/v1/manual/command-line-interface/\#Startup-file}}.
The details of the user-specified installation of \textsc{Form} and \textsc{Qgraf} are described as follows:
\begin{itemize}
    \item \textsc{Form} (v4.3.1 is recommended).
        \textsc{FORM\_jll.jl} provides a binary interface for \textsc{Form} on various platforms other than Windows.
        For Windows users, we recommend using the Windows Subsystem for Linux 2 (WSL2)\footnote{\url{https://learn.microsoft.com/en-us/windows/wsl/install}}, which provides a Linux virtual machine but with a performance comparable to the physical machine.
        To use the external \textsc{Form}, the users should set the environment variable \texttt{ENV["FORM"]} to the path of the \textsc{Form} in the \textsc{Julia} startup file as \texttt{ENV["FORM"] = "/path/to/form"}.
    \item \textsc{Qgraf} (v3.6.7 is recommended).
        \textsc{FeAmGen.jl} will download \textsc{Qgraf} from \url{http://cfif.ist.utl.pt/~paulo/d.html} and build it automatically if the \textsc{Fortran} compiler is provided by \texttt{ENV["FC"] = "/path/to/fortran-compiler"}.
        \textsc{FeAmGen.jl} will also try to use the command of \texttt{gfortran} in the system for building \textsc{Qgraf} if \texttt{ENV["FC"]} is not set.
        The users should agree with the license of \textsc{Qgraf} before using this package.
        The users can also invoke the \textsc{Qgraf} built by themselves by setting the variable \texttt{ENV["QGRAF"] = "/path/to/qgraf"}.
\end{itemize}

\paragraph{\textsc{PyCall.jl} for \textsc{Python} interface}
\label{sec:PyCall-info}
The UFO format is written in the \textsc{Python} programming language.
Therefore, the Julia package \textsc{PyCall.jl} is used in \textsc{FeAmGen.jl} to read the model files in the UFO format.
The UFO formats exported from the latest version of \textsc{FeynRules} are still written in \textsc{Python} 2, which had been stopped by \textsc{Python} developers in 2020\footnote{\url{https://peps.python.org/pep-0373/}}.
To read the UFO format written in the \textsc{Python} 2, \textsc{FeAmGen.jl} applies the following two approaches:
\begin{enumerate}
    \item (Not recommended)
        Build the \textsc{PyCall.jl} with \textsc{Python} 2.7.
        Please consult \url{https://github.com/JuliaPy/PyCall.jl.git} for more details.
    \item (Recommended)
        Convert the UFO formats from \textsc{Python} 2 format to \textsc{Python} 3 format.
        The conversion can be done directly by the tool script \texttt{2to3}\footnote{\url{https://docs.python.org/3/library/2to3.html}} provided by \textsc{Python} 3.
\end{enumerate}
%During the installation of \textsc{FeAmGen.jl}, \textsc{PyCall.jl} and its dependency \textsc{Conda.jl} may fail to build with error messages:
%\begin{center}
%    \includegraphics{code-BuildPyCallError1}
%\end{center}
%According to the information, the users may run \texttt{import Pkg; %Pkg.precompile()}, which will give the error messages in detail as
%\begin{center}
%    \includegraphics{code-BuildPyCallError2}
%\end{center}
%If so, the users could just run the following commands to rebuild %\textsc{PyCall.jl}:
%\begin{center}
%    \includegraphics{code-BuildPyCall}
%\end{center}
For more information, please consult \url{https://github.com/JuliaPy/PyCall.jl}.

% FeAmGen-Release (c) by Quan-feng WU <wuquanfeng@ihep.ac.cn> and Zhao Li <zhaoli@ihep.ac.cn>
% 
% FeAmGen-Release is licensed under a
% Creative Commons Attribution 4.0 International License.
% 
% You should have received a copy of the license along with this
% work. If not, see <http://creativecommons.org/licenses/by/4.0/>.

\section{Basic Concepts and Algorithms}\label{sec:concept}

\subsection{Diagrams and Amplitudes}

Feynman diagram is the core technique in the pQFT calculation.
In Feynman diagram technique, one can generate all the Feynman diagrams and their corresponding amplitudes for a given process up to a certain perturbation order according to the Feynman rules.

The Feynman rules are derived from the specified Lagrangian utilizing the path integral approach.
\textsc{FeynRules} is a \textsc{Mathematica} package capable of automating the generation of Feynman rules directly from the Lagrangian, which exports these rules in the UFO format for further application.

\begin{figure}[htbp]
    \centering
    \subcaptionbox{Diagram 1 (Sign: 1, Symmetry factor: 1) \label{fig:gbtW-diagram-1}}[.45\textwidth]{\includegraphics{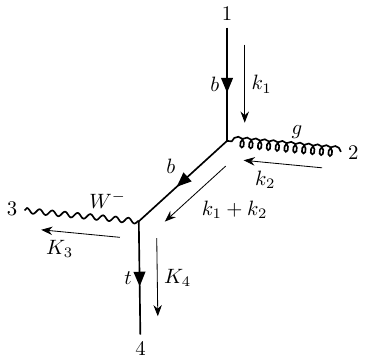}}
    \subcaptionbox{Diagram 2 (Sign: 1, Symmetry factor: 1)}[.45\textwidth]{\includegraphics{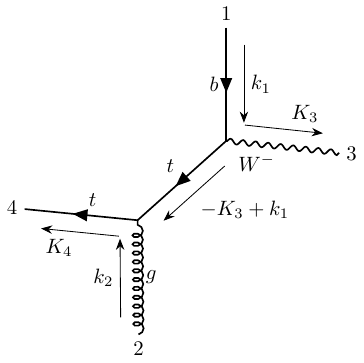}}
    \caption{
        Feynman diagrams for $g b \to t W^-$ at tree-level, generated by \textsc{FeAmGen.jl}, where $K_j$'s represent the massive external momenta and $k_j$'s represent the massless external momenta.
    }
    \label{fig:gbtW-diagrams}
\end{figure}

\textsc{FeAmGen.jl} translates UFO model file into the model file for \textsc{Qgraf}.
\textsc{Qgraf} is then used to generate the Feynman diagrams for the given process.
For visualizing the Feynman diagrams, the \LaTeX{} package \textsc{TikZ-Feynman} is used, e.g., Fig.~\ref{fig:gbtW-diagrams} shows an example of the Feynman diagrams for the process $g b \to t W^-$ at the tree level.

\textsc{FeAmGen.jl} reads the Feynman diagrams from the output of \textsc{Qgraf} and translates them into the internal data structure of \textsc{FeAmGen.jl}.
Finally, \textsc{FeAmGen.jl} generates the Feynman amplitudes according to the Feynman diagrams and the corresponding model.
Some preliminary manipulations such as color decomposition will be applied automatically to the Feynman amplitudes before the final output.
The propagators are called the internal (external) propagators with (without) the loop momenta.
The general form of Feynman amplitudes generated by \textsc{FeAmGen.jl} can be expressed as
\begin{equation}
    s \qty(\sum_{i=1}^n C_i N_i) \prod_{j=1}^{n'} \frac{1}{D_j^{\nu_j}},
    \label{eq:basic-amplitude-structure}
\end{equation}
where $s$ is the signed symmetry factor, i.e., the additional minus sign from the fermion loop multiplied by the symmetry factor;
$C_i$'s are the color factors;
$D_j$'s are the denominators of the internal propagators;
$\nu_j$'s are the exponents of $D_j$'s, respectively;
And $N_j$'s are the numerators along with relevant color factors.
The loop integration measures are not provided explicitly in \textsc{FeAmGen.jl}.
For instance, we show the color factor, the numerator expression, and the signed symmetry factor for Fig.~\ref{fig:gbtW-diagram-1} in Tab.~\ref{tab:gbtW-amplitude-1}.
Since there is no loop structure in Fig.~\ref{fig:gbtW-diagram-1}, $D_j$ and $\nu_j$ are not shown here.
The original plain output is shown as
\begin{center}
    \includegraphics{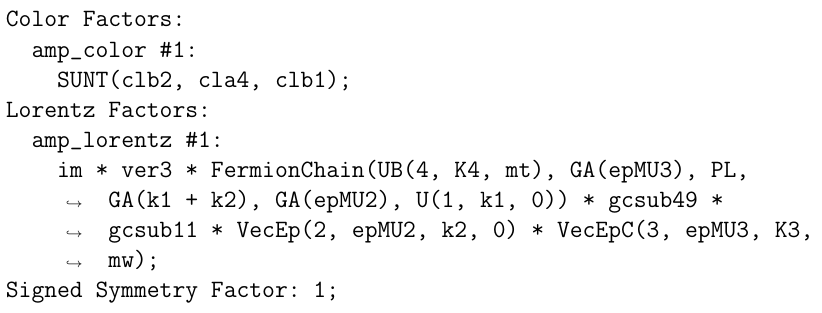}
\end{center}
Please refer to \ref{apdx:output} for the details of the expression in the output.

\begin{table}[htbp]
    \centering
    \begin{tabular}{c|l}
        item & value \\
        \hline
        $s$ & +1 \\
        $C_1$ & $T^{a_2}_{i_4, j_1}$ \\
        $N_1$ & $\mathii g \frac{\mathii e}{s_\mathrm{W} \sqrt{2}} \frac{\mathii}{(k_1 + k_2)^2 + \ieta} \qty[\bar{u}_4(K_4, m_t) \gamma_{\mu_3} P_\mathrm{L} (\slashed{k}_1 + \slashed{k}_2) \gamma_{\mu_2} u_1(k_1)] \varepsilon_2^{\mu_2}(k_2) \varepsilon_3^{*\mu_3}(K_3, m_W)$ \\
    \end{tabular}
    \caption{Amplitude of Fig.~\ref{fig:gbtW-diagram-1}.}
    \label{tab:gbtW-amplitude-1}
\end{table}

\subsection{Feynman Integrals}

For the Feynman diagrams with loop structures, the Feynman loop integrals appearing in the Feynman amplitudes are given by
\begin{equation}
    \int \qty[ \prod_{r=1}^\ell \widetilde{\dd[D]{q_r}}] T^{\mu_1 \cdots \mu_{n_t}}(q_1, \cdots, q_\ell) \prod_{j=1}^{n_\mathrm{int}} \frac{1}{\qty(p_j^2 - m_j^2 + \ieta)^{\nu_j}}
    \label{eq:general-Feynman-integral}
\end{equation}
where $\nu_j \in \mathbb{Z}$ for $1 \le j \le n_\mathrm{int}$;
$\widetilde{\dd[D]{q_r}}$ is the loop integration measure, e.g., $\widetilde{\dd[D]{q_r}} = \dd[D]{q_r} / (2 \pi)^D$ in the standard quantum field theory and $\widetilde{\dd[D]{q_r}} = \dd[D]{q_r} / \qty(\mathii \pi^{D/2})$ in most of the literature on the Feynman integrals;
$D$ is the space-time dimension, e.g., $D = 4$ in the standard quantum field theory and $D = 4 - 2 \epsilon$ usually in the dimensional regularization;
$T^{\mu_1 \cdots \mu_{n_t}}(q_1, \cdots, q_\ell)$ is the $n_t$-rank tensor constructed from the tensor products of the loop momenta $q_1, \cdots, q_\ell$;
$n_\mathrm{int}$ is the number of the internal propagators;
$m_j$ is the mass of the $j$th internal propagator;
The momentum of the $j$th internal propagator is given by
\begin{equation}
    p_j \equiv \sum_{r=1}^\ell \lambda_{j r} q_r + \sum_{s=1}^{n_\mathrm{ext} - 1} \sigma_{j s} k_s,
    \label{eq:internal-propagator-momentum}
\end{equation}
where the $\lambda_{j r}$'s and $\sigma_{j s}$'s are obtained from the Feynman diagrams via the four-momentum conservation in every vertex.
The tensor Feynman integrals are usually reduced to the scalar Feynman integrals, i.e., $T^{\mu_1 \cdots \mu_{n_t}}(q_1, \cdots, q_\ell) = 1$.
We will only focus on the scalar Feynman integrals in the following subsections.

\subsubsection{Canonical Form}\label{sec:canon}

\begin{figure}[htbp]
    \centering
    \includegraphics{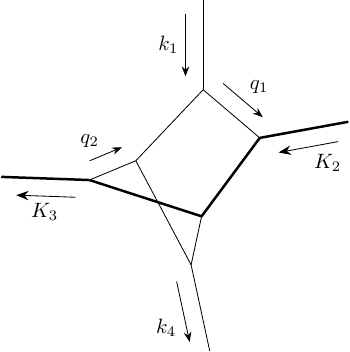}
    \caption{
        A simple example of the two-loop Feynman diagram.
        The thick lines represent the massive propagators and massive external particles with mass $m$, i.e., $K_j^2 = m^2$ here.
    }
    \label{fig:2-loop-example-1}
\end{figure}

For instance, the two-loop scalar Feynman integral of Fig.~\ref{fig:2-loop-example-1} is given as
\begin{equation}
    I_\mathrm{2-loop} = \int \widetilde{\dd[D]{q_1}} \widetilde{\dd[D]{q_2}} \frac{1}{\mathcal{D}_1 \mathcal{D}_2 \cdots \mathcal{D}_7},
\end{equation}
where
\begin{equation}
    \left\{ \begin{aligned}
        \mathcal{D}_1 & = q_1^2, \\
        \mathcal{D}_2 & = q_2^2, \\
        \mathcal{D}_3 & = (q_1 + K_2)^2 - m^2, \\
        \mathcal{D}_4 & = (q_1 - k_1)^2, \\
        \mathcal{D}_5 & = (q_2 + K_3)^2 - m^2, \\
        \mathcal{D}_6 & = (q_1 - q_2 - k_1)^2, \\
        \mathcal{D}_7 & = (q_1 - q_2 + K_2 - K_3)^2.
    \end{aligned} \right.
\end{equation}
However, this form is not unique.
The momentum shift of
\begin{equation}
    \left\{ \begin{aligned}
        q_1 & \to q_1' + k_1, \\
        q_2 & \to q_1' + q_2' - k_1
    \end{aligned} \right.
    \label{eq:momentum-shift-example}
\end{equation}
could be applied to the above integral, and the new integral is given as
\begin{equation}
    I_\mathrm{2-loop}' = \int \widetilde{\dd[D]{q_1'}} \widetilde{\dd[D]{q_2'}} \frac{1}{\mathcal{D}_1' \mathcal{D}_2' \cdots \mathcal{D}_7'},
\end{equation}
where
\begin{equation}
    \left\{ \begin{aligned}
        \mathcal{D}_1' & = (q_1' + k_1)^2, \\
        \mathcal{D}_2' & = (q_1' + q_2' - k_1)^2, \\
        \mathcal{D}_3' & = (q_1' + k_1 + K_2)^2 - m^2, \\
        \mathcal{D}_4' & = q_1'^2, \\
        \mathcal{D}_5' & = (q_1' + q_2' - k_1 + K_3)^2 - m^2, \\
        \mathcal{D}_6' & = (q_2' - k_1)^2, \\
        \mathcal{D}_7' & = (q_1' - 2 k_1 + K_2 - K_3)^2.
    \end{aligned} \right.
\end{equation}
It is obvious that $I_\mathrm{2-loop} \equiv I_\mathrm{2-loop}'$, since the Jacobian of the momentum shift of Eq.~\eqref{eq:momentum-shift-example} is $1$.
The momentum shift, therefore, introduces the redundancy in the description of the Feynman Integrals, which will hinder our further handling of the Feynman integrals, e.g. the IBP reduction.

Despite we do not implement the IBP reduction in \textsc{FeAmGen.jl}, we propose a canonical form of the Feynman integrals, which will reduce the redundancy from the momentum shift.
In short, we will choose a unique form from the equivalent Feynman integrals to represent all of them, which is the so-called canonical form in the following discussions.
The canonical form should meet the following requirements:
\begin{enumerate}
    \item The coefficients in Eq.~\eqref{eq:internal-propagator-momentum} should satisfy
        \begin{equation}
            \lambda_{jr} = \left\{ \begin{matrix}
                \delta_{jr} & j \le \ell, \\
                0 \text{ or } 1 & j > \ell,
            \end{matrix} \right.
            \qand
            \sigma_{jr} = \left\{ \begin{matrix}
                0 & j \le \ell, \\
                0 \text{ or } \pm 1 & j > \ell.
            \end{matrix} \right.
        \end{equation}
    \item The corresponding vacuum scalar integral of the Feynman integral of Eq.~\eqref{eq:general-Feynman-integral} is
        \begin{equation}
            \int \qty[\prod_{r=1}^{\ell} \widetilde{\dd^D q_r}] \prod_{j=1}^{n_\text{int}} \frac{1}{\qty(p_{j,\mathrm{vac}}^2 - m_j^2 + \ieta)^{\nu_j}}
        \end{equation}
        with
        \begin{equation}
            p_{j,\mathrm{vac}} = \sum_{r=1}^\ell \lambda_{jr} q_r.
        \end{equation}
        In the cases of up to 4 loops, the vacuum momenta $\qty{p_{j,\mathrm{vac}}}_{j=1}^{n_\mathrm{int}}$ of the canonical form of the Feynman integral should be the subset of any of the following sets:
        \begin{itemize}
            \item 1-loop: $\qty{q_1}$;
            \item 2-loop: $\qty{q_1, q_2, q_1 + q_2}$;
            \item 3-loop:
                \begin{itemize}
                    \item $\qty{q_1, q_2, q_3, q_1 + q_3, q_2 + q_3, q_1 + q_2 + q_3}$;
                    \item $\qty{q_1, q_2, q_3, q_1 + q_2, q_1 + q_3, q_2 + q_3}$;
                \end{itemize}
            \item 4-loop:
                \begin{itemize}
                    \item $\{q_1, q_2, q_3, q_4, q_1 + q_2, q_1 + q_3, q_2 + q_4, q_1 + q_2 + q_3, q_1 + q_2 + q_4, q_1 + q_2 + q_3 + q_4\}$;
                    \item $\{q_1, q_2, q_3, q_4, q_1 + q_2, q_1 + q_3, q_2 + q_4, q_3 + q_4, q_2 + q_3 + q_4, q_1 + q_2 + q_3 + q_4\}$;
                    \item $\{q_1, q_2, q_3, q_4, q_1 + q_2, q_1 + q_3, q_2 + q_3, q_2 + q_4, q_1 + q_2 + q_4, q_2 + q_3 + q_4\}$.
                \end{itemize}
        \end{itemize}
    \item In the internal propagator momenta $p_j$'s, the coefficients of the loop momenta $q$'s and external momenta $k$'s (for massless particles and $K$'s for massive particles) should be $0$ or $\pm 1$.
\end{enumerate}
Although these three requirements can remove most of the momentum shifts, there are still a large number of possibilities remaining.
We will define an order to sort all remaining possible forms of the Feynman integrals and define the one with the smallest order as the canonical form.
In the next subsection, we will describe the canonicalization algorithm in detail.

\subsubsection{Canonicalization Algorithm}

To canonicalize the Feynman integrals via the momentum shift, it should be noted that the momentum shift is applied to the loop momenta, which are the direct integration variables of the Feynman integrals.
In general, we need the momentum shift not to give an extra Jacobian, i.e.
\begin{equation}
    \abs{\det T} \equiv 1,
\end{equation}
where
\begin{equation}
    T := \pdv{\qty(q_1', \cdots, q_\ell')}{(q_1, \cdots, q_\ell)}.
\end{equation}

We only consider the momentum shifts that are just the re-determinations, where internal propagator momenta are the independent loop momenta in the corresponding Feynman diagrams.
This ensures that the Feynman amplitudes are directly related to the corresponding Feynman diagrams.
To achieve this, we first consider the momentum shift without external momenta, which can be expressed as the linear transformation of the loop momenta, i.e., the transformation matrix $\tilde{T}$ satisfies
\begin{equation}
    \tilde{q} = \tilde{T} q,
\end{equation}
where $\tilde{q} \equiv (\tilde{q}_1, \cdots, \tilde{q}_\ell)^\mathrm{T}$ and $q \equiv (q_1, \cdots, q_\ell)^\mathrm{T}$.

\begin{figure}[htbp]
    \centering
    \includegraphics{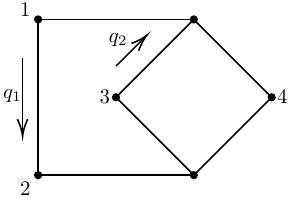}
    \caption{Internal graph of \autoref{fig:2-loop-example-1}, where the number $i$ label the vertex connecting the $i$-th external leg that has been removed.}
    \label{fig:2-loop-example-internal-graph}
\end{figure}

We introduce the \textit{internal graph} associated with the Feynman diagram, which is obtained by removing all external legs but preserving the vertices connecting external legs from the Feynman diagram.
For instance, Fig.~\ref{fig:2-loop-example-internal-graph} is the internal graph associated with Fig.~\ref{fig:2-loop-example-1}.
The internal graph is the directed graph, which is used to describe the momentum flow in the relevant Feynman diagram.
The directions of edges could be assigned arbitrarily, which will introduce an overall plus or minus to every momentum.
In Fig.~\ref{fig:2-loop-example-internal-graph}, the unique momenta are given as $\qty{q_1, q_2, q_1 - q_2}$.
Then, we could choose any two edges in the internal graph and relabel their momenta as $\pm \tilde{q}_1$ or $\pm \tilde{q}_2$.
If the chosen edges are valid, i.e., the momenta are linearly independent, then the momentum shift could be obtained by solving the linear equations $\tilde{q} = \tilde{T} q$ with $\tilde{T}_{ir} = \pm \lambda_{jr}$,
where $i$ is the index of the shifted momentum and $j$ is the index of the chosen edge.
Therefore, the momentum shift without external momenta is simply given as $q \to \tilde{T}^{-1} \tilde{q}$.

The external momenta are then associated with the loop momenta.
Following the same procedure, we choose any two propagators from the original Feynman diagram and redefine their propagator momenta as $\pm q_1'$ or $\pm q_2'$.
Obviously,
\begin{equation}
    q_i' = \tilde{q}_i \pm \sum_{m=1}^{n_\mathrm{ext}-1} \sigma_{jm} k_m,
\end{equation}
where $\tilde{q}_i$'s are the corresponding shifted loop momenta without external momenta.
Then,
\begin{equation}
    q_i' \mp \sum_{m=1}^{n_\mathrm{ext}-1} \sigma_{jm} k_m = \qty(\tilde{T} q)_i.
\end{equation}
Finally, we have the momentum shift with external momenta as
\begin{equation}
    q \to \tilde{T}^{-1} \qty(q' \mp \sigma \cdot k).
\end{equation}

The canonicalization algorithm searches all possible momentum shifts that could be derived from the original Feynman diagrams as described above.
Then, the requirements of the canonical form of the Feynman integrals will be checked.
However, an enormous number of possible momentum shifts are still allowed.
We propose an ordering to sort all possible shifted propagator momenta, and the leading one will be chosen as the canonical form.
The default ordering is defined by a two-entry tuple of integers.
Considering the case with 3 loops and 4 external legs (3 independent external momenta), the first entry is calculated by the following steps:
\begin{enumerate}
    \item For every propagator momentum, the normalization should be applied.
    This means that the coefficient of the first loop momentum $q_i$ with a non-zero coefficient in the propagator momenta should be normalized to $+1$, where the first means the smallest index of the loop momenta.
    For instance, $-q_2 + q_3 - K_1 + k_2$ should be normalized to $q_2 - q_3 + K_1 - k_2$;
    \item Because of the second and the third requirements of the canonical form, the coefficients of all loop momenta and external momenta are either $0$ or $\pm 1$.
        For every propagator momentum, we could calculate the coefficients of all loop momenta and external momenta.
        For instance, the coefficients of normalized propagator momentum $q_2 - q_3 + K_1 - k_2$ are shown as the following table:
        \begin{center}
            \begin{tabularx}{.75\textwidth}{c | >{\centering\arraybackslash}X >{\centering\arraybackslash}X >{\centering\arraybackslash}X | >{\centering\arraybackslash}X >{\centering\arraybackslash}X >{\centering\arraybackslash}X}
                \multirow{2}{*}{Momentum} & \multicolumn{3}{c|}{External momentum} & \multicolumn{3}{c}{Loop momentum} \\ \cline{2-7} & $k_3$ & $k_2$ & $K_1$ & $q_3$ & $q_2$ & $q_1$ \\
                \hline
                Coefficient & $0$ & $-1$ & $+1$ & $-1$ & $+1$ & $0$ \\
                Coefficient mod 3 & $0$ & $2$ & $1$ & $2$ & $1$ & $0$
            \end{tabularx}
        \end{center}
        % \begin{equation*}
        %     \begin{matrix}
        %         k_3 & k_2 & K_1 & q_3 & q_2 & q_1 \\
        %         0 & -1 & +1 & -1 & +1 & 0 \\
        %         0 & 2 & 1 & 2 & 1 & 0
        %     \end{matrix}
        %     \label{eq:momentum-coefficients}
        % \end{equation*}
        where we substitute $-1$ with $2$ in the last line.
        Now the the digit sequence of the coefficients of all loop momenta and external momenta is obtained as $021210$, which could be converted to a ternary number $(21210)_3 = (210)_{10}$, which is called \textit{propagator momentum number} here;
    \item Finally, the first entry of the order is the product of the propagator momentum numbers of all propagator momenta.
\end{enumerate}
The second entry is a hash number\footnote{See \url{https://docs.julialang.org/en/v1/base/base/\#Base.hash} for more details.}, which is calculated by the following steps:
\begin{enumerate}
    \item Construct a list of the propagator momentum numbers of all propagator momenta;
    \item Sort the list according to the propagator momentum numbers;
    \item Calculate the hash number of the list via the \texttt{hash} function in \textsc{Julia}.
\end{enumerate}
With the ordering defined above, the canonical form of the Feynman integrals could be obtained by sorting all possible momentum shifts with the ordering.

\subsubsection{Topologies}\label{sec:topology}

For a general form of scalar Feynman integral
\begin{equation}
    \int \qty[ \prod_{r=1}^\ell \widetilde{\dd[D]{q_r}}] \prod_{j=1}^{n_\mathrm{int}} \frac{1}{\qty(p_j^2 - m_j^2 + \ieta)^{\nu_j}},
    \label{eq:scalar-Feynman-integral}
\end{equation}
we call the set of $n_\text{int}$ denominators $\{p_j^2 - m_j^2 + \ieta\}_{j=1}^{n_\text{int}}$ as the topology of this scalar Feynman integrals.
There are, however, $N_V$($\equiv (\ell + 1) \ell / 2 + \ell (n_\text{ext} - 1)$) independent scalar products (ISPs) of loop momenta and external momenta for $\ell$-loop $n_\text{ext}$-leg Feynman diagrams.
If $n_\text{int} < N_V$, some of the ISPs cannot be represented as the linear combination of $n_\text{int}$ denominators.
Therefore, some additional propagators should be included to complete the topology.
In the comparison of two topologies, if the denominator set of the first topology is the subset of the set of the second topology, we call the first topology covered by the second topology.

The IBP identities allow one to express any scalar Feynman integral as a linear combination of a much smaller set of integrals, which are called master integrals.
Before the IBP reduction, for better efficiency one needs to minimize the number of the complete Feynman topologies that cover all the topologies of the generated Feynman amplitudes.
We implement an algorithm for it, which is sketched as follows:
\begin{enumerate}
    \item Check the covering relationships between all the topologies and drop out the topologies covered by the larger ones.
    \item Find all the suitable momentum shifts between two remaining topologies where one could cover another one after applying the suitable momentum shift.
        The topologies covered by others will also be dropped out in this step and the corresponding momentum shift will be recorded.
        Notice that the Pak algorithm \cite{Pak:2011xt, Wu:2023upw} is applied to find the suitable momentum shift.
        Please refer to \ref{apdx:Pak} for more details on the implementation of the Pak algorithm.
    \item Construct a minimal set of topologies that can cover all remaining topologies.
        In this step, the union of some topologies will be constructed, and the greedy algorithm will be applied to minimize the number of the remaining topologies.
    \item Apply step 2 again.
    \item Make the complete topologies by adding the additional denominators.
        The additional denominators are constructed from the linear combination of loop momenta in the second requirement for the canonical form of Feynman integrals, which is described in Sec.~\ref{sec:canon}.
        For example, we consider the independent external momenta $K_\mathrm{ext} = \qty{K_1, k_2, k_3}$ and the propagator momenta is a subset of the first case in the 3-loop, i.e. a subset of $Q_\mathrm{int} = \qty{q_1, q_2, q_3, q_1 + q_3, q_2 + q_3, q_1 + q_2 + q_3}$.
        Then, the ISPs in $\qty{ \qty(Q \pm K)^2 \middle| Q \in Q_\mathrm{int} \wedge K \in K_\mathrm{ext} }$ will be checked whether they can be represented as the linear combination of all the inverses of propagators.
        Otherwise, the additional propagators will be added to the denominator topologies.
    \item Check how the constructed topologies cover the original topologies.
        The momentum shift of all the original topologies will be calculated for all the constructed topologies if \texttt{recheck\_momentum\_shifts=true}.
        Otherwise, the corresponding momentum shift will be constructed from the recorded information in previous steps.
\end{enumerate}
In the steps 3 and 5, we provide two modes which can affect the implementations:
\begin{itemize}
    \item \texttt{:Canonicalization}.
        In this mode, we demand that the constructed topologies in step 3 should satisfy the requirements of the canonical form.
    \item \texttt{:PakAlgorithm}.
        In this mode, the constructed topologies in step 3 do not need to satisfy the requirements of the canonical form.
        However, step 5 is also skipped.
\end{itemize}

The bottleneck of this algorithm is searching for the momentum shift, which however can be skipped if \texttt{check\_momentum\_shifts=false}.

% FeAmGen-Release (c) by Quan-feng WU <wuquanfeng@ihep.ac.cn> and
% Zhao Li <zhaoli@ihep.ac.cn>
% 
% FeAmGen-Release is licensed under a
% Creative Commons Attribution 4.0 International License.
% 
% You should have received a copy of the license along with this
% work. If not, see <http://creativecommons.org/licenses/by/4.0/>.

\section{Usages}\label{sec:usage}

\subsection{Overview}

\begin{figure}[htbp]
    \centering
    \includegraphics{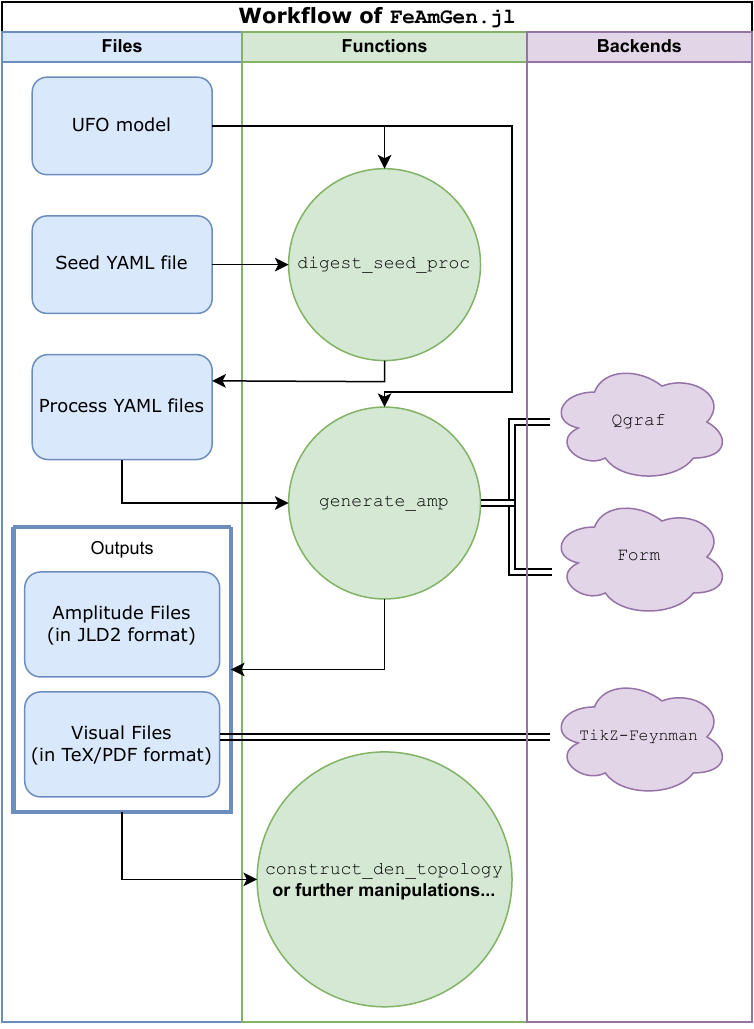}
    \caption{Workflow of \textsc{FeAmGen.jl}}
    \label{fig:FeAmGen-workflow}
\end{figure}

The sketch of \textsc{FeAmGen.jl} is shown in Fig.~\ref{fig:FeAmGen-workflow}.
The users should prepare the model file in the UFO format, which could be generated from the \textsc{Mathematica} package \textsc{FeynRules} \cite{Christensen:2009jx, Alloul:2013bka}.
Then, the configuration files should be prepared in the YAML format, which will be read by the functions of \textsc{FeAmGen.jl}.
There are two types of configuration files: the seed configuration file and the process configuration file.
The users could prepare the seed configuration file and use the function \texttt{digest\_seed\_proc} to generate the process configuration files.
In addition to generating the process configuration file by \texttt{digest\_seed\_proc}, the users can also directly edit the process configuration file according to their requirements.
After the process configuration file is prepared, the users could use the function \texttt{generate\_amp} to generate the Feynman diagrams and the corresponding amplitudes.
The details of the output will be explained in \ref{apdx:output}.

We also provide several useful functions to manipulate the generated amplitudes, e.g., the function \texttt{construct\_den\_topology} for constructing the minimal set of topologies that cover all the Feynman integrals, the function \texttt{find\_fermion\_loops} for finding the fermion loops in the Feynman diagrams, and the function \texttt{is\_planar} for checking the planarity of the Feynman diagrams.

In the following subsections, we will go into more detail about the usages of the functions and the configuration files, where the statement of using the package \textsc{FeAmGen.jl} is omitted in the following codes, which reads
\begin{center}
    \includegraphics{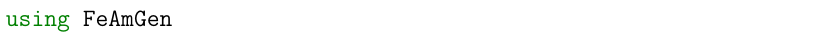}
\end{center}
The statement should be inserted before calling the functions of \textsc{FeAmGen.jl}.

\subsection{Main Functions}

\subsubsection{Digest seed (process) configuration files} 
\begin{figure}[htb]
    \centering
    \fbox{\includegraphics{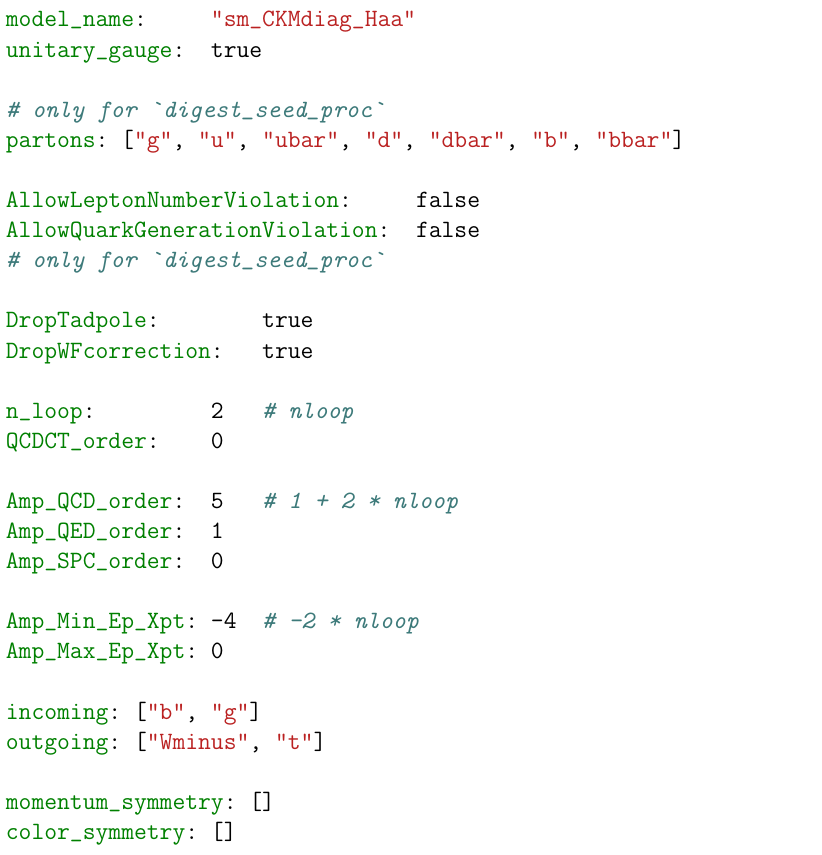}}
    \caption{Seed configuration file for the process $g b \to t W^-$ at 2-loop.}
    \label{code:gbtW-seed-configuration}
\end{figure}

The function \texttt{digest\_seed\_proc} reads the seed configuration file and generates the process configuration files.
The usage of this function is
\begin{center}
    \includegraphics{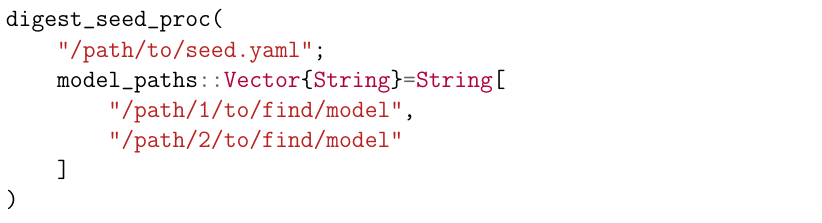}
\end{center}
The users should prepare the seed configuration file in the YAML format, e.g., Fig.~\ref{code:gbtW-seed-configuration} is the seed configuration file for the process $g b \to t W^-$.
The string \texttt{"/path/to/seed.yaml"} should be replaced by the actual path of the seed configuration file in the above sentence.
The parameter \texttt{model\_paths} is optional, which is an array of strings for specifying the directories to search for the model files.
If it is not specified, the function \texttt{digest\_seed\_proc} searches for the model files from the working directory and the model directory we provided in the package only.

In this function, the options especially for the seed configuration file will be checked, which are
\begin{itemize}
    \item \fbox{\texttt{partons}} The set of particles to be used to generate the process configuration files.
        Using \texttt{"parton"}'s in the incoming/outgoing particle list instead of the specific particle name, the function \texttt{digest\_seed\_proc} will check this option and expand every \texttt{"parton"} into several sub-processes according to the particles listed here;
    \item \fbox{\texttt{AllowLeptonNumberViolation}} (\texttt{true}/\texttt{false}) The option for allowing/forbidding the violation of the lepton number;
    \item \fbox{\texttt{AllowQuarkGenerationViolation}} (\texttt{true}/\texttt{false}) The option for allowing/forbidding the violation of the quark generation.
\end{itemize}
The last two options are only checked in the generated sub-processes.
The other options in the seed configuration file will be copied to the process configuration files directly, which are detailed as follows:
\begin{itemize}
    \item \fbox{\texttt{model\_name}} The name of the UFO model to invoke.
        It is worth mentioning that the function \texttt{digest\_seed\_proc} will search for the model files from the working directory if the location of the model file is missing.
        % We also provide several model files\footnote{The models are archived in \url{https://code.ihep.ac.cn/IHEP-Multiloop/FeAmGen_artifacts/-/tree/main/Models}, which are from \url{http://feynrules.irmp.ucl.ac.be/wiki/StandardModel} but with some minor modifications.} for searching.
    \item \fbox{\texttt{unitary\_gauge}} (\texttt{true}/\texttt{false}) The option to choose the unitary gauge or not for the internal massive gauge bosons.
    \item \fbox{\texttt{DropTadpole}} (\texttt{true}/\texttt{false}) The option for dropping the diagrams with tadpoles or not.
    \item \fbox{\texttt{DropWFcorrection}} (\texttt{true}/\texttt{false}) The option for dropping the diagrams with corrections of wave functions or not (also known as the diagrams amputated or not).
    \item \fbox{\texttt{n\_loop}} Number of loops.
    \item \fbox{\texttt{QCDCT\_order}} Order of QCD counter terms.
        If the users want to generate order-$n_\text{order}$ with order-$n_\text{CT}$ Feynman diagrams/amplitudes, the entry \texttt{n\_loop} should be set to $n_\text{order} - n_\text{CT}$ and the \texttt{QCDCT\_order} should be set to $n_\text{CT}$.
    \item \fbox{\texttt{Amp\_QCD\_order}}/\fbox{\texttt{Amp\_QED\_order}} Order of the QCD/QED coupling in the amplitudes.
    \item \fbox{\texttt{Amp\_SPC\_order}} Order of the special coupling in the amplitudes.
        The users could mark the special coupling order (SPC order) in the UFO model file for the vertices they are interested in.
        Then the users could specify the SPC order here for the diagram generation.
    \item \fbox{\texttt{Amp\_Min\_Ep\_Xpt}}/\fbox{\texttt{Amp\_Max\_Ep\_Xpt}} The minimal/maximal order of $\varepsilon$ in the series expansion of $\varepsilon$ in the amplitudes.
        Since the series expansion in $\epsilon$ is not implemented in \textsc{FeAmGen.jl}, these options are just the placeholders for the future handling of the amplitudes.
    \item \fbox{\texttt{incoming}}/\fbox{\texttt{outgoing}} The set of incoming/outgoing particles (\texttt{"parton"} is acceptable and it will be expanded by \texttt{digest\_seed\_proc}).
    \item \fbox{\texttt{momentum\_symmetry}} Symmetries between external momenta.
        This option could be needed in the evaluation of cut diagrams.
        For instance, \texttt{[K1, K3]} means $K_1 = K_3$, i.e., the replacement $K_3 \to K_1$ will be applied in the final expressions of the amplitudes.
    \item \fbox{\texttt{color\_symmetry}} Symmetries between the colors carried by external particles.
        This option could be needed in the decay or scattering process of the bound state. 
        For example, \texttt{[1, 3]} means the colors of particle 1 and particle 3 are the same, i.e. $\delta_{c_1 c_3}$ will be inserted into the color factors and then calculated automatically.
\end{itemize}

After running the function \texttt{digest\_seed\_proc}, the process configuration files will be generated and saved in the directory of the sub-directories of the working directory, which are named by the process name and the loop level.

\subsubsection{Generate amplitudes} 
\begin{figure}[htb]
    \centering
    \fbox{\includegraphics{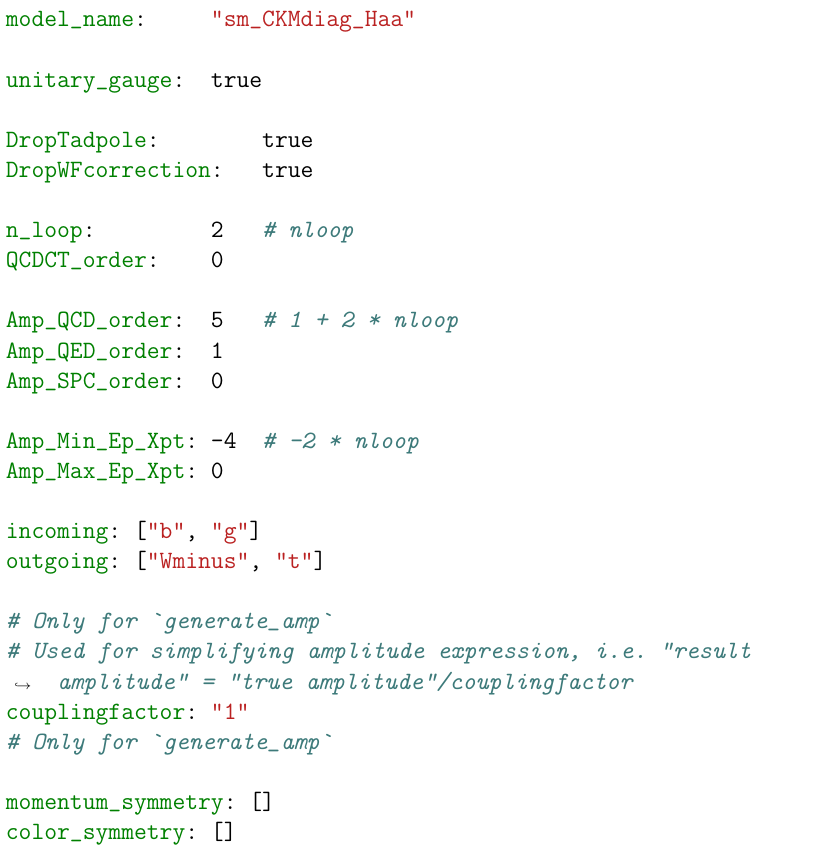}}
    \caption{Process configuration file for the process $g b \to t W^-$ at 2-loop.}
    \label{code:gbtW-process-configuration}
\end{figure}

The function \texttt{generate\_amp} reads the process configuration file and generates the Feynman diagrams and the corresponding amplitudes.
The usage of \texttt{generate\_amp} is
\begin{center}
    \includegraphics{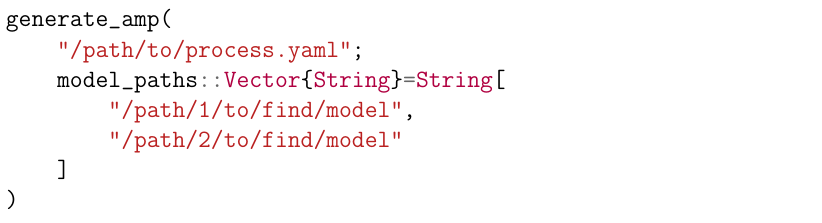}
\end{center}
Its usage is similar to the function \texttt{digest\_seed\_proc}.
The process configuration file in the YAML format is also almost structurally identical to the seed configuration file except that the three specific options for seed configuration files are removed and one more option \fbox{\texttt{couplingfactor}} is added.
For example, Fig.~\ref{code:gbtW-process-configuration} is the process configuration file for the process $g b \to t W^-$ at 2-loop.
The option \fbox{\texttt{couplingfactor}} with default value \texttt{"1"} is used to specify the coupling factor for the process, which is a string that archives the expression of the user-defined coupling factor for the process.

The external program \textsc{Qgraf} will be called by \texttt{generate\_amp} to generate the Feynman diagrams.
Then \texttt{generate\_amp} will read the output of \textsc{Qgraf} and generate the amplitudes internally.
Some color factors and the kinematic relations will be calculated by \texttt{generate\_amp} automatically, where the external CAS program \textsc{Form} will be called to do the simplification of the color factors.
The canonicalization will be applied to the amplitudes automatically, which is explained in Sec.~\ref{sec:canon}.

After running the function \texttt{generate\_amp}, the amplitude directory and the visual directory will be created as the sub-directories of the directory where the process configuration file is located.
All the amplitudes are stored in the amplitude directory in the JLD2 format\footnote{See \url{https://juliaio.github.io/JLD2.jl/stable} for more details.}, and the corresponding visual files are stored in the visual directory in the \LaTeX{} format with the package \textsc{TikZ-Feynman} \cite{Ellis:2016jkw}.

\subsubsection{Model files}\label{sec:model}

The Universal \textsc{FeynRules} Output format \cite{Degrande:2011ua} is a model file format for several automatic matrix element generation software in high energy physics.
The model files used in \textsc{FeAmGen.jl} are written in the UFO format, since it provides a flexible format archiving all the information about a theoretical model in an abstract form.
The Universal Feynman Output format (UFOv2) \cite{Darme:2023jdn} is the second version of the UFO format, which is more powerful than the first version.
However, we recommend the first version of the UFO format in \textsc{FeAmGen.jl}, which could be generated from the \textsc{Mathematica} package \textsc{FeynRules} \cite{Christensen:2009jx, Alloul:2013bka}, since the new features of the UFOv2 format are not supported in \textsc{FeAmGen.jl} yet.
We will add the support for the UFOv2 format in the future versions of \textsc{FeAmGen.jl}.

In the UFO format, the information on the particles, parameters, and vertices of the model are stored in a set of \textsc{Python2} objects.
\textsc{PyCall.jl}, the \textsc{Julia} interface to \textsc{Python}, is needed for accessing the model files in \textsc{FeAmGen.jl}.

\subsection{Other Useful Functions}\label{sec:other-functions}

\paragraph{Construct minimal set of topologies}
We provide the function \texttt{construct\_den\_topology} for constructing the minimal set of topologies that cover all the Feynman integrals.
The usage of this function is
\begin{center}
    \includegraphics{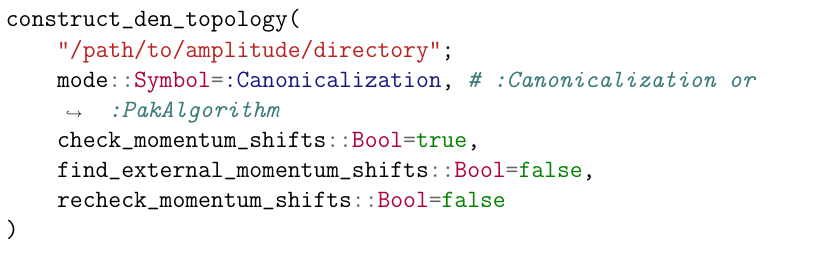}
\end{center}
There are four options for this function:
\begin{itemize}
    \item \fbox{\texttt{mode}} We provide two modes for \texttt{construct\_den\_topology}.
        \begin{itemize}
            \item \texttt{:Canonicalization} \\
                In this mode, the function \texttt{construct\_den\_topology} will try to construct the topologies that satisfy the requirements of the canonical form in Sec.~\ref{sec:canon}.
                This mode only works in the cases of 1- to 4-loop.
            \item \texttt{:PakAlgorithm} \\
                No more requirements will be added to the topologies in this mode.
                However, the complete topologies will not be constructed in this mode.
        \end{itemize}
    \item \fbox{\texttt{check\_momentum\_shifts}} (\texttt{true}/\texttt{false}) The option for checking the momentum shift or not.
        If \texttt{false}, the function \texttt{construct\_den\_topology} will not check the momentum shift, and the topologies will be constructed without momentum shift.
    \item \fbox{\texttt{find\_external\_momentum\_shifts}} (\texttt{true}/\texttt{false}) The option for finding the external momentum shift or not.
        In practice, the momentum shift including the external momentum shift is hard to find, which is the reason why this option is provided.
        We do not recommend setting this option to \texttt{true} unless the users are familiar with the details of the momentum shift.
        Please see the discussion in \ref{apdx:Pak:momentum-shifts} for more details.
        If \texttt{mode=:Canonicalization}, this option will be ignored.
    \item \fbox{\texttt{recheck\_momentum\_shifts}} (\texttt{true}/\texttt{false}) The option for rechecking the momentum shift or not.
        If \texttt{true}, the function \texttt{construct\_den\_topology} will recheck the momentum shift of the original amplitudes to the constructed topologies.
        If \texttt{false}, the function \texttt{construct\_den\_topology} will not recheck the momentum shift, and the output momentum shift is constructed from the process of constructing the topologies.
\end{itemize}
After the execution of function \texttt{construct\_den\_topology}, the minimal set of topologies will be generated and saved in the topology directory.

\begin{figure}
    \centering
    \includegraphics{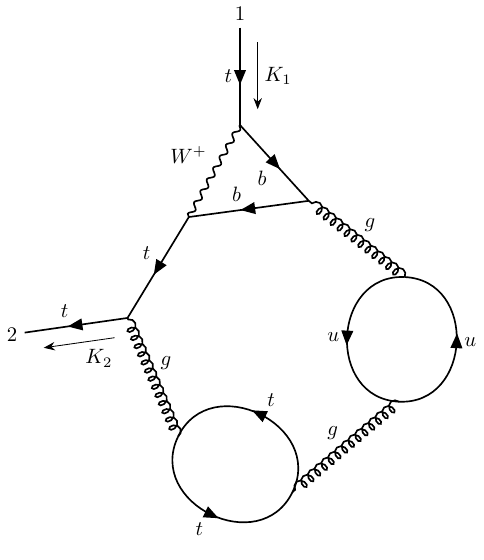}
    \caption{The 1272nd Feynman diagram for the process $t \to t W^- \to t$ at 4-loop, which contains two fermion loops.}
    \label{fig:2-fermion-loops}
\end{figure}

\paragraph{Find fermion loops}
In practice, the users may need to find the fermion loops in the Feynman diagrams.
The function \texttt{find\_fermion\_loops} is provided for this purpose, whose usage is
\begin{center}
    \includegraphics{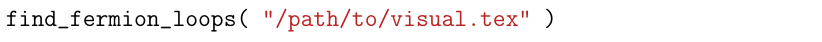}
\end{center}
The function \texttt{find\_fermion\_loops} reads the Feynman diagrams in the \LaTeX{} format generated by \textsc{FeAmGen.jl} and returns the fermion loops.
For example,
\begin{center}
    \includegraphics{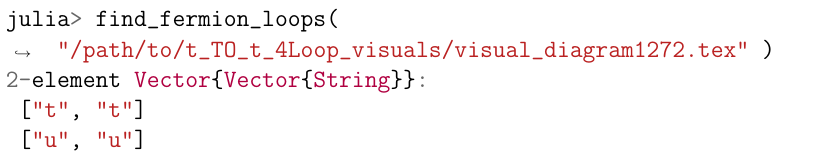}
\end{center}
The returned value is a vector of vectors of strings, where each vector of strings represents a fermion loop that is a sequence of the particle names in the loop.
The corresponding Feynman diagram is shown in Fig.~\ref{fig:2-fermion-loops}.

\paragraph{Check planarity}
The users may also want to check the planarity of the Feynman diagrams.
The function \texttt{is\_planar} is provided for this purpose, where the external program \textsc{nauty} and its \textsc{Julia} interface \textsc{nauty\_jll.jl} are used.
The usage of this function is
\begin{center}
    \includegraphics{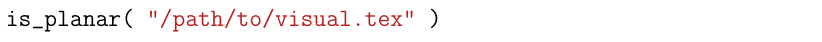}
\end{center}
The function \texttt{is\_planar} will read the Feynman diagrams in the \LaTeX{} format and return Boolean values for the planarity of the diagrams.

\subsection{Examples}\label{sec:example}

We provide several examples in the directory \href{https://code.ihep.ac.cn/IHEP-Multiloop/FeAmGen.jl/-/tree/main/test}{\texttt{FeAmGen.jl/test}}.
The users could refer to these examples for the usages of \textsc{FeAmGen.jl}.

\paragraph{Generate amplitudes}
The scripts for generating Feynman diagrams and the corresponding amplitudes are shown as follows:
\begin{itemize}
    % \item \href{https://code.ihep.ac.cn/IHEP-Multiloop/FeAmGen.jl/-/tree/main/test/ppttbar_Test.jl}{\texttt{ppttbar\_Test.jl}} for $\text{p + p} \to t +\bar{t}$ at the tree and 1-loop levels; \\
    \item \texttt{ppttbar\_Test.jl} for $p + p \to t +\bar{t}$; \\
        The incoming particles in this test are set to be \texttt{"parton"}, which will be expanded by the function \texttt{digest\_seed\_proc}.
        In subsequent steps of this test, we will only generate the amplitudes and the visuals of the sub-process $g g \to t \bar{t}$ for simplicity.
    \item \texttt{DrellYan\_Test.jl} for $u \bar{d} \to W^+$;
    \item \texttt{eeHZ\_Test.jl} for $e^+ e^- \to H Z$;
    \item \texttt{gbtw\_Test.jl} for $g b \to t W^-$;
    \item \texttt{tWb\_Test.jl} for $t \to b W^-$;
    \item \texttt{tWtW\_Test.jl} for $t W^+ \to t W^+$.
\end{itemize}

\paragraph{Construct minimal set of topologies}
For the function \texttt{construct\_den\_topology}, we provide the following examples:
\begin{itemize}
    \item \texttt{construct\_den\_topology\_Canon.jl} for the mode of \texttt{:Canonicalization};
    \item \texttt{construct\_den\_topology\_Pak.jl} for the mode of \texttt{:PakAlgorithm}.
\end{itemize}
The minimal set of topologies for the process $g b \to t W^-$ at the 2-loop level will be constructed.
Before running this script, the amplitudes for the process $g b \to t W^-$ at the 2-loop level should be prepared, which could be generated by the script \texttt{gbtw\_Test.jl}.

% FeAmGen-Release (c) by Quan-feng WU <wuquanfeng@ihep.ac.cn> and
% Zhao Li <zhaoli@ihep.ac.cn>
% 
% FeAmGen-Release is licensed under a
% Creative Commons Attribution 4.0 International License.
% 
% You should have received a copy of the license along with this
% work. If not, see <http://creativecommons.org/licenses/by/4.0/>.

\section{Conclusion}\label{sec:conclusion}

We present the Julia package \textsc{FeAmGen.jl}, a powerful and versatile Julia program that can handle the challenging task of generating Feynman diagrams and amplitudes for various quantum field theory processes.
This package is designed to meet the increasing demand for high precision theoretical predictions in particle physics, especially in the era of precise tests of the Standard Model and beyond.
It takes advantage of the UFO model format, which allows users to define arbitrary models and couplings conveniently.
In \textsc{FeAmGen.jl}, we utilize the external program \textsc{Qgraf} \cite{Nogueira:1991ex} and \textsc{Form} \cite{Vermaseren:2000nd}, the well-established external tools that can efficiently generate Feynman diagrams and simplify amplitude expressions.
The corresponding visual files will be also generated for the diagrams in the \LaTeX{} format with \textsc{TikZ-Feynman} \cite{Ellis:2016jkw}.
Moreover, \textsc{FeAmGen.jl} provides a useful function to construct topologies that cover all possible Feynman integrals of a given process at any loop level.

In this paper, the details of the algorithm and the implementation of \textsc{FeAmGen.jl} are presented.
The usages of \textsc{FeAmGen.jl} are also demonstrated by several examples.
Bug reports, feature requests, and contributions are all welcome at \url{https://code.ihep.ac.cn/IHEP-Multiloop/FeAmGen.jl.git} or \url{https://github.com/zhaoli-IHEP/FeAmGen.jl.git}.

\textit{
    Note Added: \textsc{FeAmGen.jl} has been applied in the amplitude generation of the four-loop top quark self-energy in the Standard Model \cite{Chen:2023dsi}.
    We are also working on a tiny derivative of \textsc{FeAmGen.jl} for SMEFT two-loop matching.
}

% FeAmGen-Release (c) by Quan-feng WU <wuquanfeng@ihep.ac.cn> and
% Zhao Li <zhaoli@ihep.ac.cn>
% 
% FeAmGen-Release is licensed under a
% Creative Commons Attribution 4.0 International License.
% 
% You should have received a copy of the license along with this
% work. If not, see <http://creativecommons.org/licenses/by/4.0/>.

\phantomsection
\addcontentsline{toc}{section}{Acknowledgements}
\section*{Acknowledgements}
    The authors want to thank Long-Bin Chen\orcid{0000-0002-7647-7716}, Hai-Tao Li\orcid{0000-0003-0682-2868}, Yan-Qing Ma\orcid{0000-0001-5795-3791}, Jian Wang\orcid{0000-0002-7506-3028}, Yefan Wang\orcid{0000-0002-9369-6443}, Di Wu\orcid{0000-0001-7309-574X} and Zihao Wu\orcid{0000-0003-3561-5403} for useful discussions and valuable suggestions.
    This work was supported by the National Natural Science Foundation of China under grant No. 12075251.

\appendix
\setcounter{figure}{0}
\setcounter{table}{0}
% FeAmGen-Release (c) by Quan-feng WU <wuquanfeng@ihep.ac.cn> and Zhao Li <zhaoli@ihep.ac.cn>
% 
% FeAmGen-Release is licensed under a
% Creative Commons Attribution 4.0 International License.
% 
% You should have received a copy of the license along with this
% work. If not, see <http://creativecommons.org/licenses/by/4.0/>.

\section{Outputs}\label{apdx:output}

In this appendix, we will introduce the outputs of \textsc{FeAmGen.jl} in detail.

\subsection{Process Configuration Files}\label{sec:process-config}

The output of the function \texttt{digest\_seed\_proc} is several process configuration files in the YAML format.
The function returns the array of strings, which are the paths to the process configuration files.
For instance,
\begin{center}
    \includegraphics{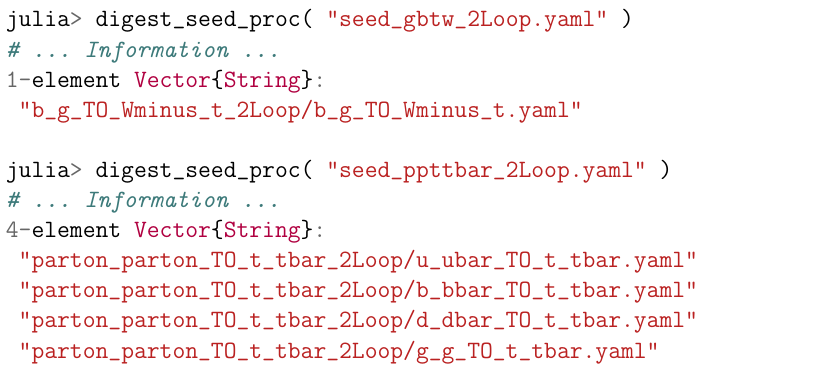}
\end{center}

\subsection{Amplitudes}\label{apdx:amp-output}

If the amplitudes are generated successfully, the function \texttt{generate\_amp} will return a two-entry tuple of strings, which are the directories that archive the amplitudes and the visual files, respectively.
For instance,
\begin{center}
    \includegraphics{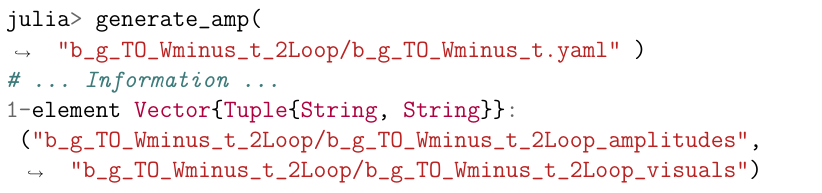}
\end{center}

\subsubsection{Amplitude Files}\label{apdx:amp-files}

The amplitudes are all archived in the directory that is the first entry of the output of \texttt{generate\_amp}.
In this directory, there are two types of files, the JLD2 files with suffix \texttt{jld2} and the UTF-8 format text files with suffix \texttt{out}, which are the human-readable text files that have the same information as the JLD2 files.
It is preferred to read the JLD2 files in the programs because the JLD2 file format is designed data interchange format for \textsc{Julia}.
The \textsc{JLD2.jl}\footnote{See \url{https://juliaio.github.io/JLD2.jl/stable} for more details.} package is required for reading the JLD2 files.

The users could read the amplitude file in the JLD2 format via
\begin{center}
    \includegraphics{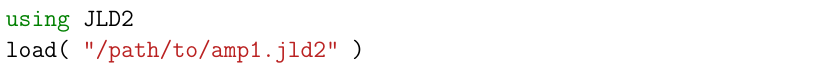}
\end{center}
where the function \texttt{load} will return a \texttt{Dict} with 18 entries.
For example, the first amplitude of the process $g b \to t W^-$ at 2-loop reads
\begin{itemize}
    \item \texttt{"Generator"}: \texttt{"FeAmGen.jl"}, the name of the generator of this amplitude.
    \item \texttt{"n\_inc"}: Number of incoming particles.
        2 for the process $g b \to t W^-$.
    \item \texttt{"n\_loop"}: Number of loops.
        2 for the process $g b \to t W^-$ at 2-loop.
    \item \texttt{"min\_ep\_xpt"}/\texttt{"max\_ep\_xpt"}: Minimal/maximal order of $\varepsilon$ in the series expansion of $\varepsilon$.
        Notice that these two entries are just the placeholders and the actual series expansion of $\varepsilon$ is not performed in \textsc{FeAmGen.jl}.
    \item \texttt{"couplingfactor"}: Coupling factor of the amplitude, which is a string of the expression of the coupling factor.
        The value of this entry is just copied from the input process configuration file.
        In general, denoting the amplitude constructed from the amplitude file as $\mathcal{A}_\text{result}$, the true amplitude for the Feynman diagram as $\mathcal{A}_\text{true}$, and the coupling factor as $c$, we have $\mathcal{A}_\text{true} = c \mathcal{A}_\text{result}$.
    \item \texttt{"ext\_mom\_list"}: The list of external momenta.
        The external momentum has the format of \texttt{K<index>} or \texttt{k<index>} for massive or massless momentum, respectively, where \texttt{<index>} is a non-negative integer representing the index of the momentum.
        For instance, the value of this entry for the process $g b \to t W^-$ reads \texttt{["k1", "k2", "K3", "K4"]} because $g$ and $b$ are massless and $t$ and $W^-$ are massive in the model we used.
    \item \texttt{"scale2\_list"}: List of mass squared scales.
        \texttt{["shat", "mw\^{}2", "ver1", "mt\^{}2"]} for the process $g b \to t W^-$, where \texttt{"shat"} means $s \equiv (k_1 + k_2)^2$, and \texttt{ver1} means $t \equiv (k_1 - K_3)^2$.
        Please refer to \ref{apdx:kin} for more information about the kinematic variables.
    \item \texttt{"loop\_den\_list"}: List of propagator denominators containing the loop momenta.
        For example, the value of this entry for the process $g b \to t W^-$ at 2-loop reads
        \begin{itemize}
            \item \texttt{"Den(q2, 0, 0)"};
            \item \texttt{"Den(K3 - k1 - k2 + q1 + q2, 0, 0)"};
            \item \texttt{"Den(K3 - k1 + q1, 0, 0)"};
            \item \texttt{"Den(q1, mt, 0)"};
            \item \texttt{"Den(q1 + q2, mt, 0)"},
        \end{itemize}
        where \texttt{"Den(P, m, ieta)"} means $\qty(P^2 - m^2 + \ieta)^{-1}$.
        Notice that the momentum starts with \texttt{q} is the loop momentum.
    \item \texttt{"loop\_den\_xpt\_list"}: List of the exponents for the corresponding propagator denominators including loop momenta. 
        \texttt{[1, 1, 1, 1, 1]} for  the process $g b \to t W^-$ at 2-loop.
        According to the information we have, the corresponding scalar integral of this amplitude is
        \begin{equation*}
            I[1,1,1,1,1] = \int \qty[\prod_{r=1}^2 \frac{\dd^D q_r}{\mathii \pi^{D / 2}}] \frac{1}{q_2^2 \qty(q_{12} + k_{123})^2 \qty(q_1 + k_{13})^2 \qty(q_1^2 - m_t^2) \qty(q_{12}^2 - m_t^2)},
        \end{equation*}
        where $k_{123} := -k_1 - k_2 + K_3$, $k_{13} := -k_1 + K_3$ and $q_{12} := q_1 + q_2$.
    \item \texttt{"kin\_relation"}: Dictionary of the kinematic relations.
        The details of the kinematic relations will be discussed in \ref{apdx:kin}.
    \item \texttt{"mom\_symmetry"}: Symmetries between the external momenta defined in the input process configuration file.
        Please refer to the \fbox{\texttt{momentum\_symmetry}} option in the process configuration file for more information.
    \item \texttt{"color\_symmetry"}: Symmetries between the colors carried by the external particles defined in the input process configuration file.
        Please refer to the \fbox{\texttt{color\_symmetry}} option in the process configuration file for more information.
    \item \texttt{"model\_parameter\_dict"}/\texttt{"model\_coupling\_dict"}: Dictionary of model parameters/couplings, which archive the expressions of the model parameters/couplings from the UFO file.
    \item \texttt{"signed\_symmetry\_factor"}: Signed symmetry factor, the product relative sign from fermion line or loop, and the symmetry factor of the Feynman diagram.
        For instance, the value of this entry for the first amplitude of the process $g b \to t W^-$ at 2-loop reads \texttt{"1"}.
    \item \texttt{"amp\_color\_list"}: List of color part expressions of the amplitude.
    \item \texttt{"amp\_lorentz\_list"}: List of Lorentz part expressions of the amplitude.
\end{itemize}
The last two entries, \texttt{"amp\_color\_list"} and \texttt{"amp\_lorentz\_list"}, which are not explained in the above list, are the most important entries of the amplitude file.
Generally, the amplitude of the quantum chromodynamics (QCD) process could be factorized into the color factors and the rest of the Lorentz covariant expressions.
Then the manipulations on the color factors and the Lorentz covariant expressions could be performed separately.

\paragraph{Color factors and Lorentz covariant expressions}
According to the Feynman rules of the QCD, the four-gluon vertex has three types of color-Lorentz structures, which reads
\begin{equation}
    \sim f^{abe} f^{cde} \qty(g_{\mu\rho} g_{\nu\sigma} - g_{\mu\sigma} g_{\nu\rho}) + f^{ace} f^{dbe} \qty(g_{\mu\sigma} g_{\rho\nu} - g_{\mu\nu} g_{\rho\sigma}) + f^{ade} f^{bce} \qty(g_{\mu\nu} g_{\sigma\rho} - g_{\mu\rho} g_{\nu\sigma}).
\end{equation}
Therefore, one four-gluon vertex will split one expression into three color factors and the corresponding Lorentz covariant expressions.
\textsc{FeAmGen.jl} will manipulate the color factors and the Lorentz covariant expressions internally and return the list of color factors and the list of Lorentz covariant expressions separately, which are one-to-one correspondence to each other.
The users could just make the inner product of the vector of color factors and the vector of Lorentz covariant expressions to get the amplitude of the QCD process.
The details of manipulations on the color factors are discussed in \ref{apdx:color}.

\paragraph{Output format of the color factors}
In the amplitude file, the color factors are archived in the \texttt{"amp\_color\_list"} entry, which is a list of color factors in the string format.
For instance, the color factor list in the first amplitude of the process $g b \to t W^-$ at 2-loop reads
\begin{center}
    \includegraphics{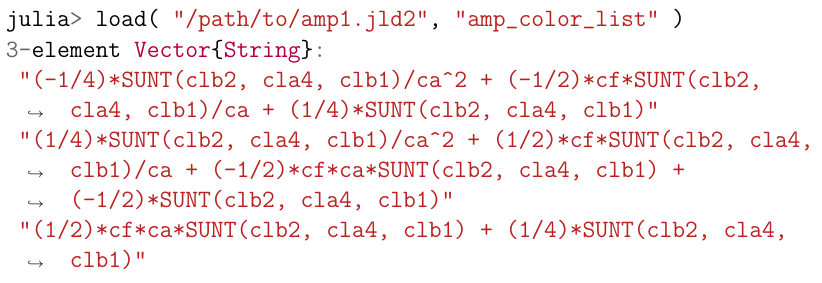}
\end{center}
There are three types of symbols in the color factors, which are detailed as follows:
\begin{itemize}
    \item \textbf{Color indices}: The color indices have the form of \texttt{cla<index>} or \texttt{clb<index>}, where \texttt{<index>} is a non-negative integer representing the index of the external particle.
        The \texttt{cla<index>} means that the particle is outgoing, and the \texttt{clb<index>} means that the particle is incoming.
        Notice that the color indices do not carry the information about the color representation of the particle, which should be constructed from the positions of the color indices in the color functions.
    \item \textbf{Color functions}: There are two types of color functions as
        \begin{itemize}
            \item \textit{Structure constant $f^{abc}$}: \\
                \texttt{SUNF(<color index a>, <color index b>, <color index c>)}
            \item \textit{Fundamental representation matrices $\qty(t_F^a)_{ij}$}: \\
                \texttt{SUNT(<color index a>, <color index i>, <color index j>)}
        \end{itemize}
    \item \textbf{Casimir invariants}: There are two Casimir invariant constants, the quadratic Casimir invariant $C_F$ (\texttt{cf}) of the fundamental representation and the quadratic Casimir invariant $C_A$ (\texttt{ca}) of the adjoint representation.
\end{itemize}

\begin{figure}[htbp]
    \centering
    \fbox{\includegraphics{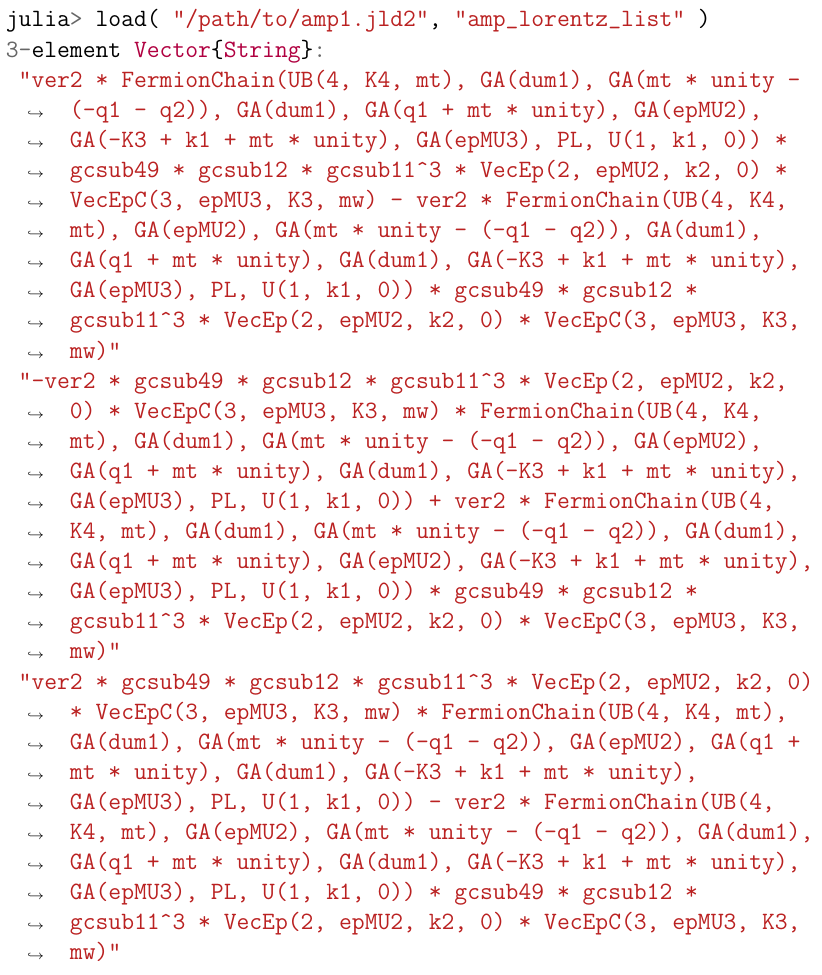}}
    \caption{Lorentz covariant expression list in the first amplitude of the process $g b \to t W^-$ at 2-loop.}
    \label{fig:amp-Lorentz-list}
\end{figure}

\paragraph{Output format of the Lorentz covariant expressions}
The Lorentz covariant expressions are archived in the \texttt{"amp\_lorentz\_list"} entry, which is a list of Lorentz covariant expressions in the string format.
For instance, the Lorentz covariant expression list in the first amplitude of the process $g b \to t W^-$ at 2-loop is shown in Fig.~\ref{fig:amp-Lorentz-list}, where some spaces are added for the readability.
There are 8 types of symbols in the Lorentz covariant expressions, which are detailed as follows:
\begin{itemize}
    \item \textbf{Model specific symbols}: The symbols defined in the UFO file, which are the model parameters and the model couplings, e.g., \texttt{gscub11} in Fig.~\ref{fig:amp-Lorentz-list}.
    \item \textbf{Kinematic symbols}: The kinematic variables, which are the scalar products of the external momenta or the Feynman denominators without loop momenta, e.g., \texttt{ver2} in Fig.~\ref{fig:amp-Lorentz-list} means \texttt{Den(k1 + k2, 0, 0)}.
        The details of the kinematic variables are discussed in \ref{apdx:kin}.
    \item \textbf{(Dummy) Lorentz indices}: The Lorentz (dummy) indices have the form of \texttt{epMU<index>} or \texttt{dum<index>} (dummy), where \texttt{<index>} is a non-negative integer representing the index of the Lorentz index.
        Notice that the Einstein summation convention is used for the repeated dummy Lorentz indices.
    \item \textbf{Momenta}: The momentum symbols have the form of \texttt{q<index>} (for loop momentum), \texttt{k<index>} (for massless external momentum) or \texttt{K<index>} (for massive external momentum), where \texttt{<index>} is a non-negative integer representing the index of the momentum.
    \item \textbf{Masses}: The mass symbols, e.g., \texttt{mt} in Fig.~\ref{fig:amp-Lorentz-list}.
    \item \textbf{Dirac-$\gamma$ matrices}: The Dirac-$\gamma$ matrices are the function \texttt{GA} and the symbols \texttt{PL}, \texttt{PR} and \texttt{GA5}, e.g., \texttt{GA(epMU2)} and \texttt{GA(dum1)} in Fig.~\ref{fig:amp-Lorentz-list} represent $\gamma^\texttt{epMU2}$ and $\gamma^\texttt{dum1}$, respectively.
        In addition, \texttt{GA(q1 + mt * unity)} in Fig.~\ref{fig:amp-Lorentz-list}, where the argument of the function \texttt{GA} contains the momenta symbol or the product of \texttt{unity} and the mass symbol, represents $\gamma_\mu q_1^\mu + m_t \cdot \mathbf{1}$.
        \texttt{GA5} represents $\gamma^5$, and \texttt{PL} and \texttt{PR} represent the left- and right-handed projectors, respectively, which are defined as
        \begin{align}
            P_L & := \frac{1}{2} \qty(\mathbf{1} - \gamma^5), \\
            P_R & := \frac{1}{2} \qty(\mathbf{1} + \gamma^5).
        \end{align}
    \item \textbf{Wave functions}: The wave function symbols.
        The spinor wave functions in Fig.~\ref{fig:amp-Lorentz-list} shows
        \begin{itemize}
            \item \texttt{UB(4, K4, mt)} means $\bar{u}_4(K_4, m_t)$;
            \item \texttt{U(1, k1, 0)} means $u_1(k_1)$.
        \end{itemize}
        The wave function symbols started with \texttt{V} or \texttt{VB} are similar to \texttt{U} or \texttt{UB}.
        For the vector bosons, the wave functions in Fig.~\ref{fig:amp-Lorentz-list} shows
        \begin{itemize}
            \item \texttt{VecEp(2, epMU2, k2, 0)} means $\varepsilon_2^\mu(k_2)$;
            \item \texttt{VecEpC(3, epMU3, K3, mw)} means $\varepsilon_3^{*\mu}(K_3, m_W)$.
        \end{itemize}
    \item \textbf{Fermion chain and trace functions}: The fermion chain function is a non-commutative multiplication of a sequence of the Dirac-$\gamma$ matrices and the wave functions, e.g., \texttt{FermionChain(UB(4, K4, mt), GA(dum1), GA(mt * unity - (-q1 - q2)), GA(dum1), GA(q1 + mt * unity), PL, U(1, k1, 0))} means
        \begin{equation*}
            \bar{u}_4(K_4, m_t) \gamma^\texttt{dum1} \qty[\gamma \cdot \qty(q_1 + q_2) + m_t \cdot \mathbf{1}] \gamma_\texttt{dum1} \qty(\gamma \cdot q_1 + m_t \cdot \mathbf{1}) P_L u_1(k_1).
        \end{equation*}
        The Fermion trace function is the trace of the non-commutative multiplication of a sequence of the Dirac-$\gamma$ matrices and the wave functions, e.g., \texttt{Trace(GA(dum1), GA(mt * unity - (-q1 - q2)), GA(dum1), GA(q1 + mt * unity), PL)} means
        \begin{equation*}
            \tr{\gamma^\texttt{dum1} \qty[\gamma \cdot \qty(q_1 + q_2) + m_t \cdot \mathbf{1}] \gamma_\texttt{dum1} \qty(\gamma \cdot q_1 + m_t \cdot \mathbf{1}) P_L}.
        \end{equation*}
        Notice that in the trace function, the spinor wave functions will be simplified by the relations of
        \begin{align}
            \sum_{s} u_s(p, m) \bar{u}_s(p, m) & = \qty(\slashed{p} + m), \\
            \sum_{s} v_s(p, m) \bar{v}_s(p, m) & = \qty(\slashed{p} - m).
        \end{align}
\end{itemize}

According to Eq.~\eqref{eq:basic-amplitude-structure}, the first amplitude of the process $g b \to t W^-$ at 2-loop could be constructed from the above information.

\subsubsection{Visual Files}

In the visual directory, the second entry of the output of \texttt{generate\_amp}, there are the visual files of the amplitudes with extension suffix \texttt{tex} and a \textsc{Julia} script named \texttt{generate\_diagram\_pdf.jl} that is used to generate the visual files.
The users could run the \textsc{Julia} script to generate the visual files in PDF format.
Notice that the \texttt{lualatex} and the \textsc{TikZ-Feynman} package are required for generating the visual files.

\subsection{Topologies}

The function \texttt{construct\_den\_topology} will return an array of the constructed Feynman topologies.
The directory that archives the topologies in the JLD2 format will be created at the directory where the amplitude directory is located, e.g.,
\begin{center}
    \includegraphics{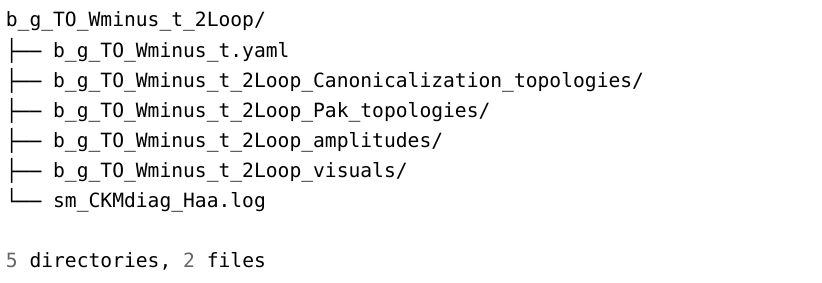}
\end{center}
There are two topology directories, which are created by different modes of the function \texttt{construct\_den\_topology}, respectively.
In every topology directory, there are the JLD2 files with suffix \texttt{jld2} and the UTF-8 format text files with suffix \texttt{out}, which are the human-readable text files that have the same information as the JLD2 files.

To access the topology file in the JLD2 format, the users could use the following code:
\begin{center}
    \includegraphics{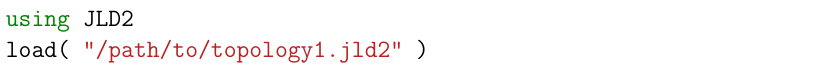}
\end{center}
Then the function \texttt{load} will return a \texttt{Dict} with 5 entries, which are
\begin{itemize}
    \item \texttt{"denominators"}: List of the propagator denominators of the topology.
        For the second topology of the process $g b \to t W^-$ at 2-loop in the mode of \texttt{:PakAlgorithm}, the value of this entry reads
        \begin{center}
            \includegraphics{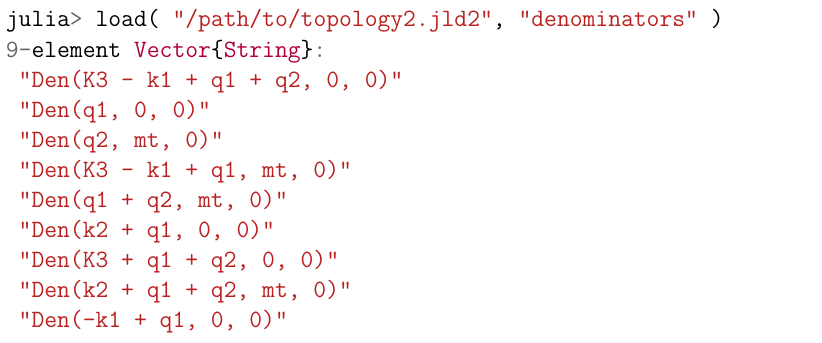}
        \end{center}
    \item \texttt{"covering\_amplitudes"} \texttt{Dict} with the paths to the amplitude files as the keys and the \texttt{Dict} of corresponding momentum shift as values.
        For the second topology of the process $g b \to t W^-$ at 2-loop in the mode of \texttt{:PakAlgorithm}, the value of this entry reads
        \begin{center}
            \includegraphics{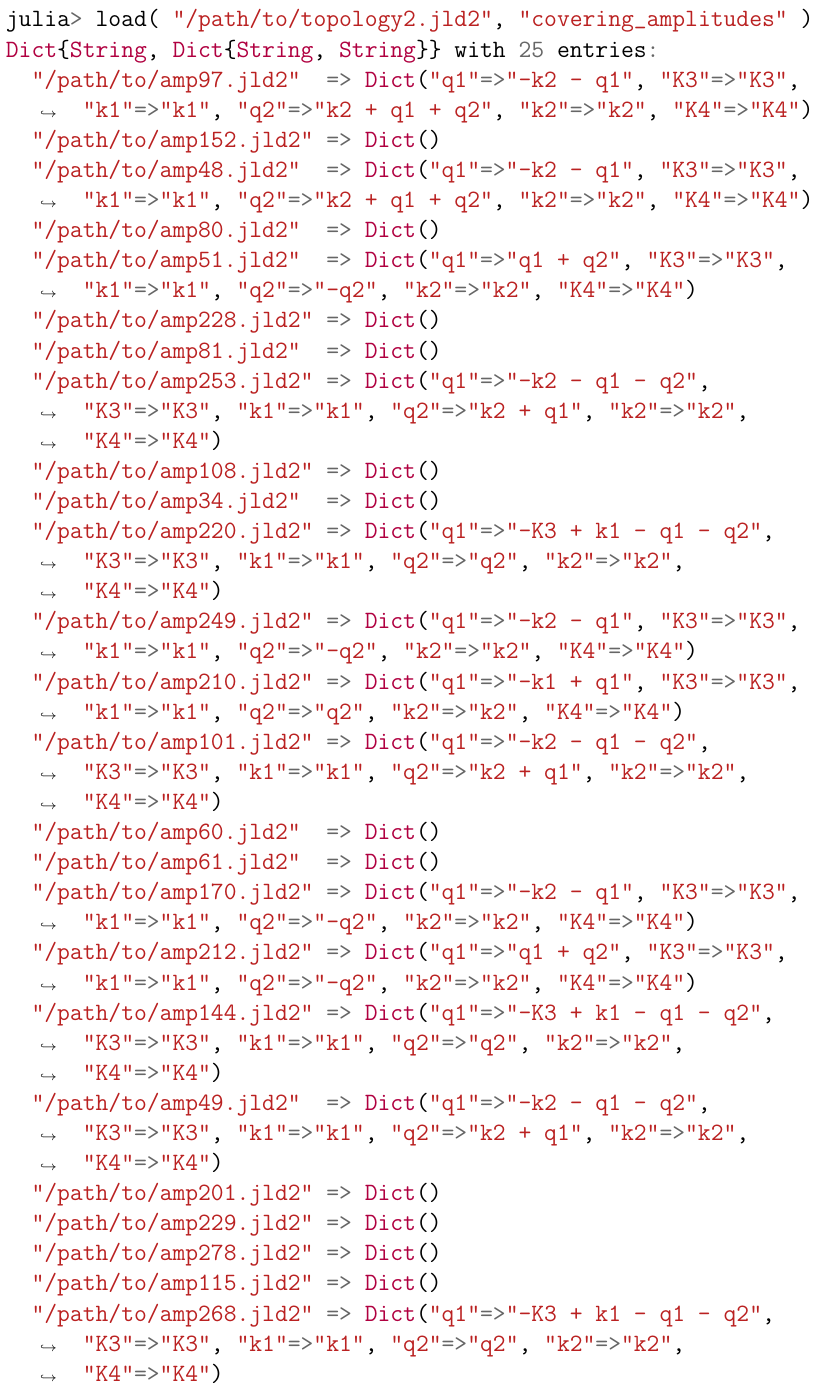}
        \end{center}
        where the empty \texttt{Dict} means that the corresponding amplitude is covered by the topology without momentum shift.
    \item \texttt{"loop\_momenta"}/\texttt{"external\_momenta"}: List of the loop/external momenta of the topology.
        For the process $g b \to t W^-$ at 2-loop, the entries \texttt{"loop\_momenta"} and \texttt{"external\_momenta"} archive \texttt{["q1", "q2"]} and \texttt{["k1", "k2", "K3", "K4"]}, respectively.
    \item \texttt{"kinematic\_relation"}: The same as the entry \texttt{"kin\_relation"} in the amplitude file, which will be discussed in \ref{apdx:kin}.
\end{itemize}

% FeAmGen-Release (c) by Quan-feng WU <wuquanfeng@ihep.ac.cn> and Zhao Li <zhaoli@ihep.ac.cn>
% 
% FeAmGen-Release is licensed under a
% Creative Commons Attribution 4.0 International License.
% 
% You should have received a copy of the license along with this
% work. If not, see <http://creativecommons.org/licenses/by/4.0/>.

\section{Kinematic Relations}\label{apdx:kin}

In this appendix, the kinematic relations used in the \textsc{FeAmGen.jl} are introduced, which are also exported to the amplitude files and topology files.
\textsc{FeAmGen.jl} only supports the decay process of $1 \to n - 1$ and the scattering process of $2 \to n - 2$ that are usually considered.

\subsection{Mandelstam Variables}

In the following discussions, all the external momenta are supposed to be outgoing.
For a process with $n$ external particles, all the momenta will be denoted by $k_i$ with $i = 1, 2, \cdots, n$.
For the incoming momenta, the extra minus sign should be multiplied.

For $n < 4$ outgoing external particles, we have
\begin{itemize}
    \item $1 \to 1$: $k_1 \cdot k_2 \equiv m_1^2$, where $k_1 \equiv k_2$ and $m_1 \equiv m_2$ are obvious;
    \item $1 \to 2$: $k_1 \cdot k_2 \equiv \frac{1}{2} \qty(m_1^2 + m_2^2 - m_3^2)$;
    \item $2 \to 1$: $k_1 \cdot k_2 \equiv \frac{1}{2} \qty(m_3^2 - m_1^2 - m_2^2)$.
\end{itemize}
For $n \ge 4$ outgoing external particles, we define the Mandelstam variables as
\begin{equation}
    \begin{aligned}
        s_{ij} & := (k_i + k_j)^2 \\
        & = k_i^2 + k_j^2 + 2 k_i \cdot k_j \\
        & = m_i^2 + m_j^2 + s'_{ij},
    \end{aligned}
    \label{eq:Mandelstam-variables}
\end{equation}
where the reduced Mandelstam variables are defined as
\begin{equation}
    s'_{ij} := 2 k_i \cdot k_j.
    \label{eq:reduced-Mandelstam-variables}
\end{equation}
There are total $n (n - 1) / 2$ reduced Mandelstam variables for $n$ external particles.
From the momentum conservation of $n$ external particles, we have
\begin{equation}
    \sum_{j=1}^n s_{i j} \equiv 0, \qfor i = 1, 2, \cdots, n.
\end{equation}
Therefore, only $n (n-1) / 2 - n = n (n-3) / 2$ reduced Mandelstam variables are independent.
To get the independent reduced Mandelstam variables, we removed the following $n$ reduced Mandelstam variables:
\begin{equation}
    s'_{1n}, \quad \cdots, \quad s'_{n-1, n}, \qand s'_{n-2, n-1}.
    \label{eq:reduced-Mandelstam-relation}
\end{equation}
Finally, the independent reduced Mandelstam variables are
\begin{equation}
    \begin{matrix}
        s'_{12}, & \cdots, & s'_{1, n-1}, \\
        s'_{23}, & \cdots, & s'_{2, n-1}, \\
        \vdots & \vdots & \vdots \\
        s'_{n-3, n-2}, & & s'_{n-3, n-1}.        
    \end{matrix}
\end{equation}
And the removed reduced Mandelstam variables are obtained by solving the equations of \eqref{eq:reduced-Mandelstam-relation}.

In \textsc{FeAmGen.jl}, we use the symbols of \texttt{shat} and \texttt{ver<index>}'s to denote the Mandelstam variables defined in \eqref{eq:Mandelstam-variables}.
For instance, the case of $n = 6$ is shown in Tab.~\ref{tab:Madelstam-variables-n6}.

\begin{table}[htb]
    \centering
    \begin{tabular}{c | c c c c c c}
        & $k_1$ & $k_2$ & $k_3$ & $k_4$ & $k_5$ & $k_6$ \\
        \hline
        $k_1$ & $m_1^2$ & $s_{12}$/\texttt{shat} & $s_{13}$/\texttt{ver1} & $s_{14}/$\texttt{ver2} & $s_{15}/$\texttt{ver3} & --- \\
        $k_2$ & --- & $m_2^2$ & $s_{23}$/\texttt{ver4} & $s_{24}$/\texttt{ver5} & $s_{25}$/\texttt{ver6} & --- \\
        $k_3$ & --- & --- & $m_3^2$ & $s_{34}$/\texttt{ver7} & $s_{35}$/\texttt{ver8} & --- \\
        $k_4$ & --- & --- & --- & $m_4^2$ & --- & --- \\
        $k_5$ & --- & --- & --- & --- & $m_5^2$ & --- \\
        $k_6$ & --- & --- & --- & --- & --- & $m_6^2$
    \end{tabular}
    \caption{The table of independent Mandelstam variables/symbols for $n = 6$.}
    \label{tab:Madelstam-variables-n6}
\end{table}

\subsection{Feynman Denominators without Loop Momenta}

In \textsc{FeAmGen.jl}, the Feynman denominators without loop momenta will also be denoted by the symbols \texttt{ver<index>}'s.
For the process $g b \to t W^-$ at 2-loop, the Feynman denominators without loop momenta in all the generated Feynman diagrams are shown as follows:
\begin{enumerate}
    \item \texttt{ver2}: \texttt{Den(k1 + k2, 0, 0)}, \texttt{Den(K3 + K4, 0, 0)};
    \item \texttt{ver3}: \texttt{Den(k1 - K3, mt, 0)}, \texttt{Den(k2 - K4, mt, 0)}.
\end{enumerate}
Notice that \texttt{ver1}\footnote{For the $2 \to 2$ process, there are only two independent Mandelstam variables which are $s$/\texttt{shat} and $t$/\texttt{ver1}.} is occupied by the Mandelstam variable which is defined in the previous discussion.
One can easily verify that \texttt{Den(K3 + K4, 0, 0) = Den(k1 + k2, 0, 0)} and \texttt{Den(k1 - K3, mt, 0) = Den(k2 - K4, mt, 0)} by using the momentum conservation $k_1 + k_2 = K_3 + K_4$.
We have also proposed an algorithm to sort them, which is sketched as follows:
\begin{enumerate}
    \item Collect all the Feynman denominators without loop momenta in all the generated Feynman diagrams;
    \item Apply the momentum conservation to remove the last external momentum in each Feynman denominator, e.g., $K_4 \to k_1 + k_2 - K_3$ in the above example;
    \item Normalize the propagator momentum of each Feynman denominator, which means the coefficient of the first external momentum $k_i$ with a non-zero coefficient in the propagator momenta should be normalized to $+1$, where the first means the smallest index of the external momenta.
        For instance, $-k_1 + K_3$ should be normalized to $k_1 - K_3$;
    \item Remove the duplicated Feynman denominators;
    \item Calculate the propagator momentum number of each Feynman denominator, which is defined in Sec.~\ref{sec:canon};
    \item Sort the Feynman denominators according to the tuple with three entries: the propagator momentum number, the string form of the second entry of the \texttt{Den} function, and the string form of the third entry of the \texttt{Den} function;
\end{enumerate}
Finally, after the symbols \texttt{ver<index>}'s that are assigned to the Mandelstam variables, we can continue to assign the symbols \texttt{ver<index>}'s to the Feynman denominators without loop momenta in this order.

% FeAmGen-Release (c) by Quan-feng WU <wuquanfeng@ihep.ac.cn> and Zhao Li <zhaoli@ihep.ac.cn>
% 
% FeAmGen-Release is licensed under a
% Creative Commons Attribution 4.0 International License.
% 
% You should have received a copy of the license along with this
% work. If not, see <http://creativecommons.org/licenses/by/4.0/>.

\section{Color Algebra}\label{apdx:color}

Color algebra is a very important part of the amplitude calculation in the QCD, which is indeed the Lie algebra $\mathfrak{su}(3)$.
\textsc{FeAmGen.jl} will make some color algebra manipulations when generating the amplitudes, where the external CAS program \textsc{Form} is required.
Before we go into the details of the color algebra manipulations, the conventions of the color algebra in \textsc{FeAmGen.jl} are introduced.
The Lie algebra $\mathfrak{su}(3)$ satisfies the Lie brackets as
\begin{equation}
    \qty[t^a, t^b] = \mathii f^{abc} t^c,
\end{equation}
where $f^{abc}$'s are the structure constants.
In the linear representation $R$ of $\mathfrak{su}(3)$, the same Lie brackets read
\begin{equation}
    \qty[t_R^a, t_R^b] = \mathrm{i} f^{abc} t_R^c,
\end{equation}
where $t_R^a$'s are the $D(R) \times D(R)$ squared representation matrices of the generators.
The Casimir invariants $T(R)$ and $C(R)$ (sometimes we use $T_R$ and $C_R$ for simplicity) of the representation $R$ are defined by
\begin{equation}
    \tr[t_R^a t_R^b] \equiv T(R) \delta^{ab}, \quad t_R^a t_R^a \equiv C(R) \cdot \mathbf{1},
\end{equation}
where $\mathbf{1}$ is $D(R) \times D(R)$ identity matrix.
The Casimir invariants $T(R)$ and $C(R)$ are related by
\begin{equation}
    T(R) D(A) \equiv C(R) D(R),
\end{equation}
where $A$ means the adjoint representation, which is given by
\begin{equation}
    (t_A^a)^{bc} \equiv -\mathrm{i} f^{abc}.
\end{equation}
In the QCD, the fundamental representation $F$ of $\mathfrak{su}_3$ is important, whose dimension is $D(F) = N$ (sometimes we use $N$ instead of $F$ for clarity).
Therefore, we have the following Casimir constants:
\begin{align}
    T(F) & = \frac{1}{2}, \\
    C(F) & = \frac{N^2 - 1}{2 N}, \\
    T(A) & = C(A) = N.
\end{align}
Notice that $C_A \equiv N$, so we have
\begin{align}
    C_A & \equiv N, \\
    C_F & \equiv \frac{C_A^2 - 1}{2 C_A}.
\end{align}
We also have other useful formulae as follows
\begin{align}
	t_R^a t_R^b t_R^a & \equiv \qty[C_R - \frac{1}{2} C_A] t_R^b, \\
	f^{acd} f^{bcd} & \equiv C_A \delta^{ab}, \label{eq:rewite-f-1} \\
	f^{abc} t_R^b t_R^c & \equiv \frac{\mathii}{2} C_A t_R^a, \label{eq:rewrite-f-2} \\
	(t_F^a)_{ij} (t_F^a)_{k\ell} & \equiv T_F \qty(\delta_{i\ell} \delta_{kj} - \frac{1}{C_A} \delta_{ij} \delta_{k\ell}). \label{eq:fundamental-relation}
\end{align}
With the above discussions, the color algebra manipulations in \textsc{FeAmGen.jl} are sketched as follows:
\begin{enumerate}
    \item Rewrite the color structure constants in terms of the products of the fundamental representation matrices of the generators by Eqs.~\eqref{eq:rewite-f-1} and \eqref{eq:rewrite-f-2}.
    \item Apply Eq.~\eqref{eq:fundamental-relation} repeatedly until no further simplification could be performed.
    \item Contract Krönecker-$\delta$'s and simplify the color factors.
\end{enumerate}
More details about the color algebra manipulations can be found in the \textsc{Form} source code in \textsc{FeAmGen.jl}.

% FeAmGen-Release (c) by Quan-feng WU <wuquanfeng@ihep.ac.cn> and Zhao Li <zhaoli@ihep.ac.cn>
% 
% FeAmGen-Release is licensed under a
% Creative Commons Attribution 4.0 International License.
% 
% You should have received a copy of the license along with this
% work. If not, see <http://creativecommons.org/licenses/by/4.0/>.

\section{Pak's Algorithm}\label{apdx:Pak}

In this appendix, Pak's algorithm is introduced, which is implemented in the \textsc{Julia} package \textsc{PakAlgorithm.jl} that is used in the \textsc{FeAmGen.jl}.
\textsc{PakAlgorithm.jl} is also developed by the authors of the \textsc{FeAmGen.jl}, which is avaliable at \url{https://code.ihep.ac.cn/IHEP-Multiloop/PakAlgorithm.jl.git}.
Any suggestions and contributions are welcome.

\subsection{Graph Polynomials}

%Before we go to the details of the Pak's algorithm, the graph polynomials should be introduced first.
For the general scalar Feynman integral of Eq.~\eqref{eq:scalar-Feynman-integral}, we define
\begin{equation}
    \sum_{j = 1}^{n_\text{int}} x_j \qty(p_j^2 - m_j^2 + \ieta) \equiv \sum_{r = 1}^\ell \sum_{s = 1}^\ell M_{r s} \qty(q_r \cdot q_s) + 2 \sum_{r = 1}^\ell q_r \cdot v_r + J,
    \label{eq:auxiliary-polynomial}
\end{equation}
where $M$ is a $\ell \times \ell$ matrix with scalar entries, $v$ is an $\ell$-vector with momentum vectors as entries and $J$ is a scalar.
Then the Symanzik polynomials $\mathcal{U}$ and $\mathcal{F}$ defined as
\begin{align}
    \mathcal{U} & := \det M, \\
    \mathcal{F} & := \qty(\sum_{r, s = 1}^\ell \qty[M^{-1}]_{r s} v_r \cdot v_s - J) \frac{\mathcal{U}}{\mu^2},
\end{align}
where $\mu^2$ is a mass-squared scale that is introduced to make the polynomials dimensionless.
In the original paper of Pak's algorithm \cite{Pak:2011xt}, the product $\mathcal{U} \mathcal{F}$ is used.
However, Ref.~\cite{Wu:2023upw} suggests the sum $\mathcal{G} = \mathcal{U} + \mathcal{F}$, which is called Lee-Pomeransky polynomial, for the implementation of the Pak's algorithm.
We will use the Lee-Pomeransky polynomial $\mathcal{G}$ instead of the product $\mathcal{U} \mathcal{F}$ for convenience.

\subsection{Canonical Lee-Pomeransky Polynomial and Pak's Algorithm}

In the auxiliary polynomial of Eq.~\eqref{eq:auxiliary-polynomial}, one could relabel the indices of $x_i$'s, which will not change the value of the original Feynman integral.
However, the relabeling will change the graph polynomials $\mathcal{U}$ and $\mathcal{F}$, hence the Lee-Pomeransky polynomial $\mathcal{G}$.
To avoid the relabeling, the naïve approach is to find a canonical permutation of $x_i$'s, and the corresponding canonical Lee-Pomeransky polynomial $\mathcal{G}$, which is unique for a given Feynman integral.
Pak proposed an algorithm to find the canonical permutation in Ref.~\cite{Pak:2011xt}, which is the so-called Pak's algorithm.

We implement Pak's algorithm in the \textsc{Julia} package \textsc{PakAlgorithm.jl} with some slight modifications, where the algorithm we have implemented is as follows:
\begin{enumerate}
    \item Construct the Lee-Pomeransky polynomial $\mathcal{G}$, which is the multivariate polynomial of $x_i$'s with $m$ monomials.
    \item Turn the Lee-Pomeransky polynomial $\mathcal{G}$ into a matrix with rows corresponding to every monomials.
        For every monomial, the first entry is the hash number of the coefficient and the $(i+1)$-th entry of the row is the exponent of $x_i$ in the monomial.
    \item Make $n_\text{int}$ new sub-matrices by extracting the first column and the $(i + 1)$-th column from the original matrix, where $i = 1, 2, \cdots, n_\text{int}$.
    \item For every sub-matrix, sort the rows lexicographically.
    \item Compare the final columns of the sorted sub-matrices to find the lexicographically largest one.
        Without loss of generality, we assume the lexicographically largest column is $x_{j_1}$, then the canonical permutation should contain $x_{j_1} \to x_1$.
    \item Make $(n_\text{int} - 1)$ new sub-matrices by extracting the first column, the $x_{j_1}$-th column and the $(i + 1)$-th column from the original matrix, where $i = 1, 2, \cdots, n_\text{int}$ but $i \neq j_1$, again for every sub-matrix, sort the rows lexicographically, and compare the final columns to find the lexicographically largest one (denoted as $x_{j_2}$).
        The canonical permutation now are constructed as $x_{j_1} \to x_1$ and $x_{j_2} \to x_2$.
    \item Repeat the above steps until all the $x_i$'s are relabeled.
\end{enumerate}
According to the constructed canonical permutation, the canonical Lee-Pomeransky polynomial $\mathcal{G}$ is obtained.

\subsection{Momentum Shift between Equivalent Feynman Topologies}\label{apdx:Pak:momentum-shifts}

We can now check if two Feynman topologies are equivalent according to their canonical Lee-Pomeransky polynomials.
If two Feynman topologies are equivalent, the canonical Lee-Pomeransky polynomials should be the same.
And the corresponding canonical permutations are also constructed.
We can then check the momentum shift between these Feynman topologies according to the constructed canonical permutations with the ansatz of
\begin{equation}
    \left\{
        \begin{aligned}
            q_i & \to \sum_{j = 1}^{\ell} A_{i j} q_j + \sum_{j = 1}^{n_\text{ext}} B_{i j} k_j, \qfor i = 1, 2, \cdots, \ell, \\
            k_m & \to \sum_{n = 1}^{n_\text{ext}} C_{m n} k_n, \qfor m = 1, 2, \cdots, n_\text{ext},
        \end{aligned}
    \right.
\end{equation}
where $\abs{\det A} \equiv 1$ for keeping the integration measure invariant and the transformation matrix $C$ should guarantee the Gram determinant of the external momenta invariant.
If the Feynman topology $T_1$ is equivalent to the sub-topology $T_2'$ of the other Feynman topology $T_2$, the momentum shift for $T_1$ could also be obtained by the above ansatz via checking the canonical permutations of $T_1$ and $T_2'$.

\textsc{PakAlgorithm.jl} provides the function \texttt{is\_Pak\_equivalent} to check if two Feynman topologies are equivalent and the function \texttt{is\_Pak\_covering} to check if one Feynman topology is covering the other one.
\texttt{find\_Pak\_momentum\_shifts} is also provided to find the momentum shift between two Feynman topologies which are equivalent or one is covering the other one.
These functions are used in the function \texttt{construct\_den\_topology} of \textsc{FeAmGen.jl}, which are detailed in Sec.~\ref{sec:topology} and Sec.~\ref{sec:other-functions}.

\bibliographystyle{elsarticle-num}
\bibliography{from-inspirehep}

\end{document}